\documentclass[12pt, draftclsnofoot, onecolumn]{IEEEtran}
\usepackage{amssymb}
\usepackage{cite}
\usepackage{graphicx}
\usepackage{array}
\usepackage{multirow}
\usepackage{amsmath}
\usepackage{stfloats}
\usepackage{graphicx}
\usepackage{subfigure}
\usepackage{tabularx}
\usepackage{epsfig,epsf,balance,cite}
\usepackage[caption=false,font=normalsize,labelfont=sf,textfont=sf]{subfig}
\usepackage{setspace}
\usepackage{bm}
\usepackage{textcomp}
\usepackage{algorithmic}
\usepackage{algorithm}
\usepackage{subfigure}
\usepackage{caption}
\usepackage{graphicx}
\usepackage{setspace}
\usepackage{url}
\usepackage{booktabs}
\usepackage{verbatim}
\usepackage{graphicx}
\usepackage{cite}
\usepackage{amssymb}
\usepackage{color}
\usepackage{xpatch}
\definecolor{shadecolor}{RGB}{0,0,255}
\definecolor{blue}{RGB}{0,0,255}

\newtheorem{theorem}{Theorem}

\makeatletter

\ExplSyntaxOn
\cs_new:Npn \bibColoredItems #1#2
{
	\clist_map_inline:nn {#2} { \cs_new:cpn {bib@colored@##1} {#1} } 
}
\ExplSyntaxOff

\newcommand\bib@setcolor[1]{%
	\ifcsname bib@colored@#1\endcsname
	\expanded{\noexpand\color{\csname bib@colored@#1\endcsname}}%
	\else
	\normalcolor
	\fi
}

\xpatchcmd\@bibitem
{\item}
{\bib@setcolor{#1}\item}
{}{\fail}

\xpatchcmd\@lbibitem
{\item}
{\bib@setcolor{#2}\item}
{}{\fail}
\makeatother

\begin{document}
	
\title{Novel Synchronization Scheme for Cooperative ISAC Systems}
	
\author{Qihao Peng, Hong Ren, Zhendong Peng, Cunhua Pan,~\IEEEmembership{Senior Member,~IEEE}, \\
	Maged Elkashlan,~\IEEEmembership{Senior Member,~IEEE},\\ Dongming Wang, ~\IEEEmembership{Senior Member,~IEEE}, Jiangzhou Wang,~\IEEEmembership{Fellow,~IEEE}, and Xiaohu You,~\IEEEmembership{Fellow,~IEEE}.
	\thanks{Q. Peng is affiliated with 5G and 6G Innovation Centre, Institute for Communication Systems (ICS) of University of Surrey, Guildford, GU2 7XH, UK.  (e-mail: q.peng@surrey.ac.uk). Zhendong Peng is with the Department of Electrical and Computer Engineering, The University of British Columbia, Vancouver, BC V6T 1Z4, Canada (e-mail: zhendongpeng@ece.ubc.ca). M. Elkashlan is with the School of Electronic Engineering and Computer Science at Queen Mary University of London, U.K. (e-mail: maged.elkashlan@qmul.ac.uk). Jiangzhou Wang is with the School of Engineering, University of Kent, CT2 7NT Canterbury, U.K. (e-mail: j.z.wang@kent.ac.uk). H. Ren, C. Pan, Dongming Wang, and Xiaohu You are with National Mobile Communications Research Laboratory, Southeast University, Nanjing, China. (e-mail:\{hren,cpan,wangdm,xhyu\}@seu.edu.cn).  (Corresponding author: Hong Ren and Cunhua Pan)} 
}

\maketitle

\begin{abstract}
	Carrier frequency and timing synchronization play the fundamental roles in cooperative integrating communication and sensing (ISAC). To mitigate the effects of synchronization error, this paper develops a novel synchronization scheme in cell-free massive multiple-input multiple-output (mMIMO) systems. First, we characterize the impacts of pilot contamination on synchronization performance, i.e., Cramer-Rao bound (CRB). Furthermore, a maximum likelihood algorithm is presented to estimate the CFO and TO among the pairing APs. Then, to minimize the sum of CRBs, we devise a synchronization strategy based on a pilot-sharing scheme by jointly optimizing the cluster classification, synchronization overhead, and pilot-sharing scheme, while simultaneously considering the overhead and each AP's synchronization requirements. To solve this NP-hard problem, we simplify it into two sub-problems, namely cluster classification problem and the pilot sharing problem. To strike a balance between synchronization performance and overhead, we first classify the clusters by using the K-means algorithm, and propose a criteria to find a good set of master APs. Then, the pilot-sharing scheme is obtained by using the swap-matching operations. Simulation results validate the accuracy of our derivations and demonstrate the effectiveness of the proposed scheme over the benchmark schemes.
\end{abstract}	

\begin{IEEEkeywords}
	Cooperative ISAC, synchronization, cell-free massive MIMO, pilot assignment, graph theory.
\end{IEEEkeywords}
	
\section{Introduction}
Enabling cooperative integrating communication and sensing (ISAC) in cell-free massive multiple-input multiple-output (mMIMO) systems have emerged as a promising technology for the sixth generation (6G) systems to simultaneously meet the ever-growing demand on data rate, sensing, and decoding error probability \cite{interdonato2019ubiquitous,peng2022resource,8999605,9737357}. By geographically deploying a multitude of distributed access points (APs) and eliminating the cell boundary, the distributed APs can seamlessly cooperate to mitigate inter-cell interference, and then perform cooperative transmission and sensing, which consequently improves the spectral efficiency and sensing accuracy. However, as each AP is based on its own local oscillator (LO), the carrier frequency offset (CFO) and timing offset (TO) among distributed APs leads to asynchronous reception, which severely deteriorates the cooperative transmission and sensing performance \cite{yan2019asynchronous,10769985}. Therefore, to reap the promising gains provided by multiple points cooperation, it is crucial to address the fundamental synchronization challenges in the distributed architectures.

One method is to connect all distributed APs by physical wires, i.e., coaxial cables or optical fibers. Though this method seems logically straightforward, it has economic and technical
challenges for practical implementation owing to substantial costs in terms of installation and maintenance. Another solution is to deploy a global positioning system (GPS) disciplined oscillator at each AP as an accurate time source for individual devices \cite{sallouha2019localization}, which is also typically expensive and impractical.

To address the above issues, extensive contributions have been devoted to the realization of wireless synchronization \cite{balan2013airsync,abari2015airshare,alemdar2021rfclock,rashid2022frequency,matsuura2022synchronization,mghabghab2022adaptive,rogalin2014scalable,feng2018frequency,ganesan2023beamsync}. Generally, the widely used methods can be divided into two categories, namely reference signal-based synchronization \cite{balan2013airsync,abari2015airshare,alemdar2021rfclock,matsuura2022synchronization,rashid2022frequency,mghabghab2022adaptive} and pilot-based synchronization \cite{rogalin2014scalable,ganesan2023beamsync}. 
For the synchronization based on reference signal, a master AP transmits an out-of-band reference signal, while slave APs can synchronize the frequency and phase by tracking the reference signal \cite{balan2013airsync}. This approach still causes various errors due to different LOs. To mitigate the effect of LO, the authors of \cite{abari2015airshare} proposed an AirShare scheme where the distributed receivers adjust LO to generate a frequency equal to the difference of two tones. Then, this strategy was further investigated in \cite{rashid2022frequency} and also adopted in distributed microwave power transfer systems \cite{matsuura2022synchronization}. However, these synchronization schemes based on reference signals utilize extra bandwidth and circuits, which causes lower spectral efficiency and higher power consumption. To tackle this issue, the authors of \cite{rogalin2014scalable} proposed a pilot-pair synchronization strategy where the slave APs estimate the CFO and TO by receiving the burst pilot from the master APs. Then, the design of optimal pairs was studied in \cite{feng2018frequency}. Inspired by the pilot-pair scheme, a similar strategy was applied to the distributed RadioWeaves infrastructure \cite{ganesan2023beamsync}. However, these synchronization schemes based on pairing pilot \cite{rogalin2014scalable,feng2018frequency,ganesan2023beamsync} did not analyze the impact of pilot allocation on the estimation performance.

Currently, various pilot allocation schemes have been adopted in cell-free mMIMO systems \cite{attarifar2018random,liu2019tabu,buzzi2020pilot,liu2020graph,zeng2021pilot,peng2023resource}. Particularly, it was demonstrated that pilot contamination is related to the distance among
users that reuse the pilots \cite{attarifar2018random}. Then, to minimize pilot contamination, the system performance can be enhanced by using the Tabu-search-based algorithm \cite{liu2019tabu} and the Hungarian algorithm \cite{buzzi2020pilot}. Furthermore, to further reduce the pilot overhead, the pilot allocation relying on graph theory was proposed in \cite{liu2020graph}. To meet the diverse requirements of users, pilot sharing schemes based on graphic theory were investigated in \cite{zeng2021pilot,peng2023resource}. However, the relationship between synchronization performance and pilot allocation is still unexplored. Furthermore, how to design the optimal synchronization scheme with limited overhead is still an open problem.

To fill this gap,  this paper aims to reveal the relationship between synchronization error and pilot allocation scheme and balance the synchronization overhead and performance, by jointly optimizing the cluster and pilot-sharing strategy. Our contributions are summarized as follows:

\begin{enumerate}
	\item By considering the impact of pilot-sharing schemes on estimation errors, we analyze a more practical scenario where several slave APs with uncertain CFOs and TOs share a common pilot sequence. Then, a new CRB for pilot-pair estimation is derived while considering the pilot contamination. Based on the given CRB, the impacts of pilot sequence and pilot allocation strategy on synchronization performance are analyzed. Finally, a maximum likelihood (ML) algorithm is provided to estimate the CFO and TO among two pairing APs.
	\item To obtain an achievable lower bound of synchronization in cell-free mMIMO systems, we aim to minimize the sum of CRBs by jointly optimizing the cluster classification and pilot allocation strategy while taking APs' requirements and limited overhead into consideration. To solve this NP-hard problem, we simplify it into two sub-problems, including the cluster classification and the pilot sharing scheme. For the first sub-problem, the requirements on CFO and TO are transformed into the desired signal-to-interference-and-noise ratio (SINR), based on which we derive the maximum distance in the cluster. Then, the clusters are classified by iteratively performing the K-means algorithm, and a good set of master APs is found by using the proposed criteria. To solve the second sub-problem, a pilot sharing scheme based on the swap-matching operations is proposed to maximize the synchronization performance with limited overhead.
	\item Our simulation results validate the accuracy of our derivations but also demonstrate that increasing the synchronous pilot length and reusing distance can significantly increase the synchronization performance. Furthermore, classifying clusters can shorten the distance between master APs and slave APs, which can enhance the path gain and significantly improve the estimation performance. Finally, our simulation results also confirm its superiority over other pilot-pair algorithms.
\end{enumerate}

The remainder of this paper is organized as follows. In Section II, the system model is provided, and the CRB and ML estimations are given, respectively. By optimizing the cluster classification, pilot length and pilot sharing scheme, the sum of CRBs is minimized in Section III. Our simulation results are presented in Section IV. Finally, our conclusions are drawn in Section V.
	
\section{System Model}
\begin{figure}
	\centering
	\includegraphics[width=4.5 in]{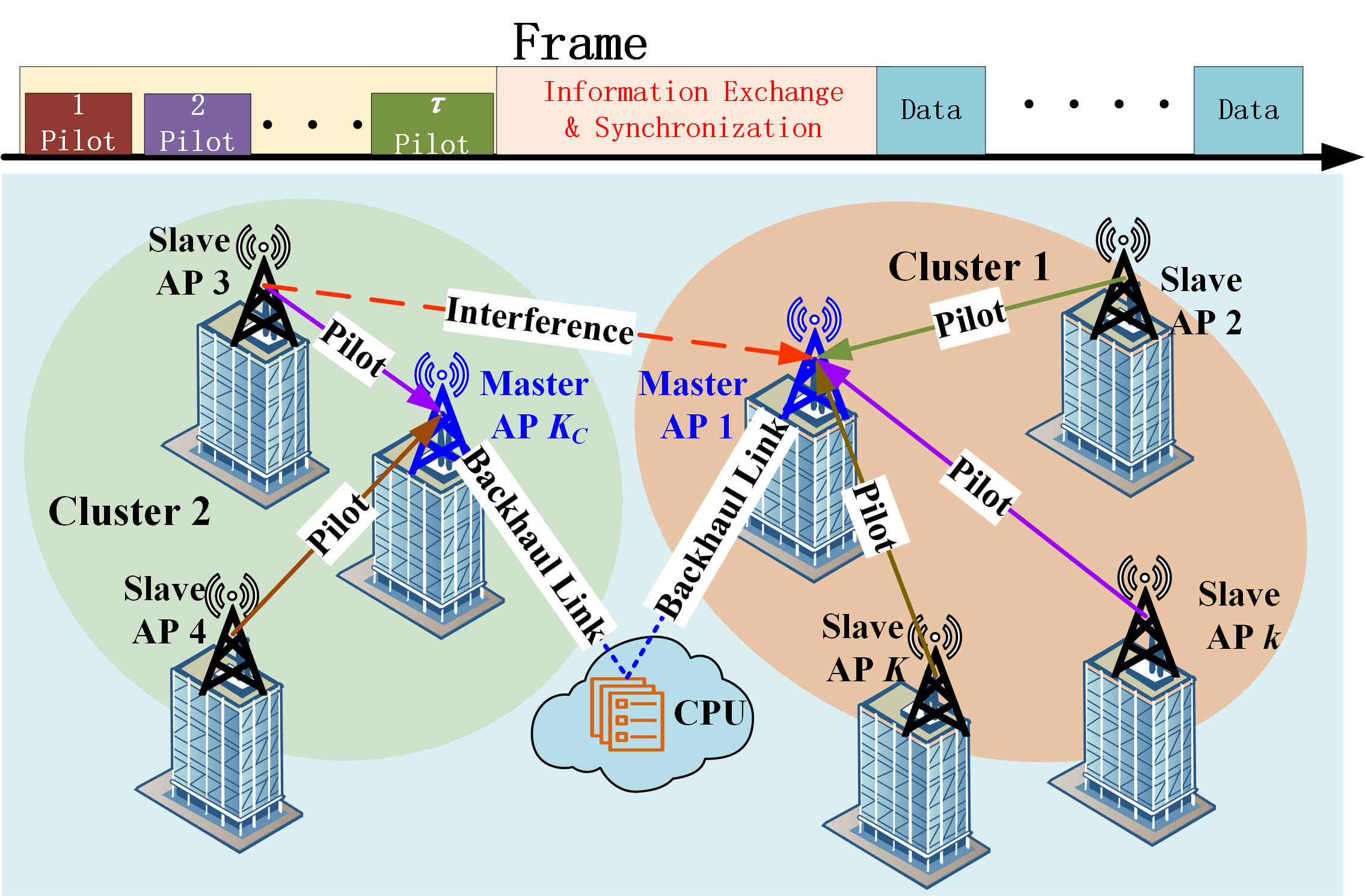}
	\caption{Illustration of pilot-pair synchronization in cell-free mMIMO systems.}
	\label{system}
\end{figure}
Since each AP has a unique oscillator and the synchronization is non-ideal, the very small residual CFO and TO will cause slow relative drift among different APs, which has a severe impact on cooperative communication and sensing performance. To prevent these errors from accumulating over time, the APs need to exchange the information and perform synchronization frequently. However, owing to the excessive number of APs, it is impractical for all APs to exchange the synchronization information with the central processing unit (CPU) directly by wireless channels, which leads to high overhead and low efficiency. 

As illustrated in Fig. \ref{system}, we consider a cooperative ISAC scenario enabled by a cell-free mMIMO system. To solve the scalability of this system, $K$ APs can be divided into $K_C$ non-overlapping clusters, denoted as ${\mathcal C}_k$, $\forall k \in \{1,2,\cdots,K_{C}\}$, and ${\cal C}_k \cap {\cal C}_{k'} = \emptyset $, if $k \ne k'$. Furthermore, we assume that each cluster has only one master AP and several slave APs. We denote the master AP in the $k$-th cluster as $ j_{k} $ and the set of slave APs in the $k$-th cluster as ${\mathcal S}_k$ with ${\cal S}_k \cup \{j_{k}\} = {\cal C}_k$ and ${\cal S}_k \cap \{j_{k}\} = \emptyset $. To synchronize all APs with low overhead, one AP may share a common pilot sequence with some specific slave APs from other clusters. Firstly, slave APs first transmit the burst pilot to master APs. Then, master APs can simultaneously receive the pilot from the multiple clusters, and then exchange the estimated CFO and TO parameters with the CPU. Finally, the CPU performs the synchronization operation by sending commands or performing compensation.


We consider a MIMO-OFDM system. Furthermore, by assuming that the antennas at each AP are based on a common local oscillator or each AP will perform self-synchronization by relying on Argos algorithm \cite{Argos}, the antennas of each AP will be perfectly synchronized. However, owing to the slow relative drift among different APs and imperfect synchronization, we denote the CFO and TO between the $i$-th AP ($i \in \{ {\mathcal S}_1 \cup \cdots \cup {\mathcal S}_{K_C} \}$) and the $j_k$-th AP as $\Delta f_{i,j_k}$ and $\Delta \tau_{i,j_k}$, respectively.  Based on the above discussions, the CFO and TO between the transmit antennas of the $i$-th AP and the $j_k$-th AP's receive antennas are the same. Therefore, for ease of derivation, we apply the received signal based on the single-antenna model for analysis.

We consider joint TO and CFO pilot-aided estimation for a transmitter-receiver pair, where the channel is multi-path time-invariant with additive white Gaussian noise (AWGN), with unknown path coefficients and multiple path delays. Generally, the number of the channel taps $L_{i,j_k}$ between the $i$-th slave AP and the $j_k$-th master AP does not exceed the length of the cyclic prefix (CP). The multi-path channel impulse response from the $i$-th slave AP to the $j_k$-th master AP can be written as
\begin{equation}
	\label{ijthchannel}
 {h_{i,j_k}}\left( t \right) = \sum\limits_{l = 0}^{{L _{i,j_k}} - 1} {\alpha _{i,j_k}^l\sqrt {{\beta _{i,j_k}}} \psi^l_{i,j_k} \delta \left( {t - \varsigma _{i,j_k}^l} \right)} ,
\end{equation}
where $\alpha _{i,j_k}^l$ and $\varsigma_{i,j_k}^l$, $\forall l \in \{0,1,2,\cdots,L_{i,j_k}-1\}$, are the path gain coefficient and path delay of the $l$-th channel tap, respectively. $\beta_{i,j_k}$ is the distance-dependent path loss, $\psi^l_{i,j_k} \in \mathcal{CN}(0,1)$ denotes the small-scale fading factor, and $\delta(t)$ is the impulse response. 

In the phase of network setup, the slave APs transmit the pilot sequence for synchronization based on a pilot burst transmission. Then, the $j_k$-th master AP receives time-domain baseband signal from several slave APs that share the common pilot sequence, which is given by
\begin{equation}
	\setlength\abovedisplayskip{5pt}
	\setlength\belowdisplayskip{5pt}
	\label{jth}
	\begin{split}
		{y_{j_k}}\left( t \right) 
		= &\! \!\sum\limits_{i \in {{\cal P}_p}} h_{i,j_k}(t) \otimes s_p(t){e^{j2\pi \Delta {f_{i,j_k}}t}} + {w_{i,j_k}}\left( t \right) \\
		= & \sum\limits_{i \in {{\cal P}_p}}\! {\sum\limits_{l = 0}^{{L}_{i,j_k} - 1}} {h_{i,j_k}^l{s_p}\!\left( {t \!-\! {\Delta \tau _{i,j_k}} \!- \!\varsigma _{i,j_k}^l} \right)}   \times {e^{j2\pi \Delta {f_{i,j_k}}(t-{\Delta \tau _{i,j_k}})}} + {w_{i,j_k}}\left( t \right), 
	\end{split}
\end{equation}
where ${\cal P}_p$ is the set of slave APs that share the $p$-th pilot sequence, $h^l_{i,j_k} = \alpha_{i,j_k}^l\sqrt {{\beta _{i,j_k}}} \psi^l_{i,j_k}$ is the $l$-th path coefficient, $\otimes$ is the convolution operator, and $w_{i,j_k} \in \mathcal{CN}(0,\sigma^2)$ represents the additive noise. $s_p (t) = \frac{P_t}{{\sqrt {{T_s}} }}\sum\limits_{n = 0}^{{N} - 1} {s_p^n\prod \left( {\frac{t}{{{T_s}}} - n} \right)} $ denotes the $p$-th pilot sequence, $s_p^n$ is the time-domain chip symbol with ${\bf s}_p=\{s_p^0, s_p^1,...,s_p^{N -1}\} $, $T_s$ is the chip interval, $P_t$ is the transmission power, and $N$ is the length of pilot sequence. $\prod(t)$ denotes the rectangular pulse, which can be expressed as
\begin{equation}
	\setlength\abovedisplayskip{5pt}
	\setlength\belowdisplayskip{5pt}
	\prod \left( t \right) = \left\{ {\begin{array}{*{20}{c}}
			{1,}&{0 \le t \le 1}\\
			{0,}&{{\rm{otherwise}}}.
	\end{array}} \right.
\end{equation}
Then, the receiver performs chip-matched filtering, and sampling per one chip by substituting $t = mT_s$, $m \in \{0,\cdots, M-1\}$. For ease of expression, we denote $\nabla (t) = \prod(t) \otimes  \prod(t)$ as the convolution of two rectangular pulses. The received discrete-time signal is given by \cite{rogalin2014scalable}
\begin{equation}
	\setlength\abovedisplayskip{5pt}
	\setlength\belowdisplayskip{5pt}
	\label{synmodel}
	\begin{split}
		{y_{j_k}}\left[ m \right] = \sum\limits_{i \in {{\cal P}_p}} \Big [ \sum\limits_{l = 0}^{{L_{i,j_k}} - 1} h_{i,j_k}^l\sum\limits_{n = 0}^{N - 1} s_p^n  {\nabla \big( {m - n - \frac{{\Delta {\tau _{i,j_k}} + \varsigma _{i,j_k}^l}}{{{T_s}}}} \big)}   \Big]  {e^{j2\pi \Delta {f_{i,j_k}}({T_s}m - \Delta {\tau _{i,j_k}})}} + {w_{i,j_k}}\left[ m \right]. 
	\end{split}
\end{equation}

By using a coarse frame synchronization protocol and compensation based on previous estimation, the receiver can mitigate the effects of the phase rotation ${e^{-j2\pi \Delta {f_{i,j_k}} \Delta {\tau _{i,j_k}}}}$ by considering it as a part of phase offset and has a coarse knowledge of the frame timing such that it expects to find the pilot burst in a given interval including the guard time \cite{rogalin2014scalable}. The master AP collects $M$ samples such that the sampled interval of duration $M T_s$ contains the pilot sequence.

Next, we collect the $M$ received samples into an $M \times 1$ vector. The $j_k$-th master AP's received signal is defined as ${\bf{y}}_{j_k} = {\left[ {y_{j_k}\left[ 0 \right],y_{j_k}\left[ 1 \right], \ldots ,y_{j_k}\left[ {{M} - 1} \right]} \right]^T}$. Then, the CFO matrix is defined as
\begin{equation}
	\setlength\abovedisplayskip{5pt}
	\setlength\belowdisplayskip{5pt}
	\label{CFOmatrix}
	\begin{split}
		{\bf{F}}\left( {{\Delta f_{i,{j_k}}}} \right) \!=\! {\rm{diag}}\left( {1,{e^{j2\pi {\Delta f_{i,{j_k}}}{T_s}}}, \!\cdots\! ,{e^{j2\pi \left( {M \!-\! 1} \right){\Delta f_{i,{j_k}}}{T_s}}}} \right).
	\end{split}
\end{equation}
The $M \times \left( {M - N + 1} \right)$ matrix of the $p$-th pilot sequence is given by
\begin{equation}
	\setlength\abovedisplayskip{5pt}
	\setlength\belowdisplayskip{5pt}
	\label{psequence}
	{{\bf{S}}_p} = \left[ {\begin{array}{*{20}{c}}
			{s_p^0}&0& \ldots &0\\
			{s_p^1}&{s_p^0}&{}& \vdots \\
			\vdots &{s_p^1}&{}&{s_p^0}\\
			{s_p^{{N} - 1}}& \vdots & \ddots &{s_p^1}\\
			0&{s_p^{N - 1}}&{}& \vdots \\
			\vdots & \vdots &{}&{}\\
			0&0& \cdots &{s_p^{{N} - 1}}
	\end{array}} \right].
\end{equation}
The $[n,l]$-th element of $ (M-N+1) \times L_{i,{j_k}}$ convolution matrix ${\bf U}(\Delta \tau_{i,{j_k}})$ is defined as
\begin{equation}
	\setlength\abovedisplayskip{5pt}
	\setlength\belowdisplayskip{5pt}
	[{\bf U}(\Delta \tau_{i,{j_k}})]_{n,l} =  \nabla \Big({n - 1 - \frac{{{\Delta \tau _{i,{j_k}}}}}{{{T_s}}} - \frac{{\varsigma _{i,{j_k}}^l}}{{{T_s}}}}\Big).
\end{equation} 
Based on the above definitions, we arrive at the vector observation model as follows
\begin{equation}
	\setlength\abovedisplayskip{5pt}
	\setlength\belowdisplayskip{5pt}
	\label{model}
	{{\bf{y}}_{j_k}} = \sum\limits_{i \in {{\cal P}_p}} {{\bf{F}}\left( {{\Delta f_{i,{j_k}}}} \right){{\bf{S}}_p}{\bf{U}}\left( {{\Delta \tau_{i,{j_k}}}} \right){{\bf{h}}_{i,{j_k}}}}  + {{\bf{w}}_{i,{j_k}}},
\end{equation}
where ${{\bf{w}}_{i,{j_k}}} = {\left[ {w_{i,{j_k}}\left[ 0 \right],w_{i,{j_k}}\left[ 1 \right], \ldots ,w_{i,{j_k}}\left[ {{M} - 1} \right]} \right]^T}$ is the vector of noise, and ${{\bf{h}}_{i,{j_k}}}$ is a vector that collects the coefficients of ${L_{i,{j_k}}}$ paths, denoted as ${{\bf{h}}_{i,{j_k}}} = {\left[ {h_{i,{j_k}}^0,h_{i,{j_k}}^1, \cdots ,h_{i,{j_k}}^{{L_{i,{j_k}}} - 1}} \right]^T}$. The derivations of (8) can be found in Appendix \ref{Proof}.


\section{Cramer-Rao bound with Pilot Contamination}
Based on the given model, we jointly estimate the synchronization parameters and derive the CRB for the CFO and TO. However, it is challenging to obtain CFO and TO from all slave APs, and thus the master APs can only estimate the synchronization parameters of the slave AP in the common cluster by treating the pilot contamination from other clusters as noise. The received signal from the $j_k$-th master AP can be rewritten as
\begin{equation}
	\label{remodel}
	\begin{split}
		{{\bf{y}}_{j_k}} & = {\bf{F}}\left( {{\Delta f_{i_{j_k}^p,j_k}}} \right){{\bf{S}}_p}{\bf{U}}\left( {{\Delta \tau _{i_{j_k}^p,j_k}}} \right){{\bf{h}}_{i_{j_k}^p,j_k}} + \underbrace{\sum\limits_{i \in {\left\{ {{{\cal P}_p}\backslash i_{j_k}^p} \right\}}} {{\bf{F}}\left( {{\Delta f_{i,j_k}}} \right){{\bf{S}}_p}{\bf{U}}\left( {{\Delta \tau_{i,j_k}}} \right){{\bf{h}}_{i,j_k}}}  + {{\bf{w}}_{i,j_k}}}_{{ \boldsymbol{\bar w}}_{i,j_k}},
	\end{split}
\end{equation}
where $i_{j_k}^p$ is the slave AP using the $p$-th pilot sequence in cluster ${\cal C}_k$ \footnote{Similar to the frequency multiplexing, it is impossible to reuse the common pilot sequence in the same cluster, which is discussed in Section IV.}.

\subsection{Cramer-Rao Bound}
In this subsection, we derive the CRB for the synchronization parameters with pilot contamination by assuming that the path delays are known \footnote{Due to the fact that the $i_{j_k}^p$-th slave AP and the ${j_k}$-th master AP are in the same cluster, the distances from the $i_{j_k}^p$-th slave AP to the ${j_k}$-th master AP is closer than that of the slave APs in other clusters and the variation of multi-path delay is relatively slow compared to inter-cluster ones. Furthermore, since the synchronization is performed periodically in a short time interval and all APs are deployed in high locations with relatively stable propagation paths, these factors lead to the known path delays.}. However, owing to the unknown TO and CFO from the slave APs that share the common pilot sequence, it is challenging to derive the exact distribution. By assuming that the pilot contamination follows the complex Gaussian distribution \footnote{Owing to no cell boundary and a large number of APs, a pilot sequence needs to be shared lots of times, and thus the pilot contamination can be approximated to follow a multivariate Gaussian distribution, by using the central limit theorem \cite{6487427}.}, we adopt the well-known conclusion for the Fisher matrix of the Gaussian probability density function to derive CRB. 

Similar to \cite{rogalin2014scalable}, we denote $\boldsymbol{\Xi}  = \text{var}\{{ \boldsymbol{\bar w}}_{i,{j_k}}\}$ as the variance of ${ \boldsymbol{\bar w}}_{i,{j_k}}$ and define $\boldsymbol{m}(\boldsymbol{\theta}) = {\bf{F}}\left( {{\Delta f_{i_{j_k}^p,{j_k}}}} \right){{\bf{S}}_p}{\bf{U}}\left( {{\Delta \tau _{i_{j_k}^p,{j_k}}}} \right){{\bf{h}}_{i_{j_k}^p,{j_k}}}$, where the elements of $\boldsymbol{\theta}$ are given by
\begin{equation}
	\label{vector}
	\theta_1 = \frac{{{\Delta \tau _{i_{j_k}^p,{j_k}}}}}{{{T_s}}}, \theta_2 = {\Delta f_{i_{j_k}^p,{j_k}}} T_s, \theta_{2l+3} = \text{Re}\{h_{i_{j_k}^p,{j_k}}^l\}, \theta_{2l+4} = \text{Im}\{h_{i_{j_k}^p,{j_k}}^l\}, \forall l \in \{0,..., {{L}_{i_{j_k}^p,{j_k}} - 1}\}. 
\end{equation}
Then, the $[n,m]$-th element of the Fisher matrix can be expressed as
\begin{equation}
	\label{Fisher}
	[{\bf J}]_{n,m}(\boldsymbol{\theta}) = 2 \text{Re} {\Big\{(\frac{\partial \boldsymbol{m}(\boldsymbol{\theta})}{\partial \theta_n})^H \boldsymbol{\Xi }^{-1}  (\frac{\partial \boldsymbol{m}(\boldsymbol{\theta})}{\partial \theta_m}) \Big\} }.
\end{equation}

Based on the above discussion, we then derive the variance of pilot contamination by assuming that the TO and CFO follow the uniform distribution of $\frac{{{\Delta \tau _{i,{j_k}}}}}{{{T_s}}} \in [0, \eta  ]$, $\eta \in {\mathbb N}^+$, \footnote{Since the coarse synchronization is performed periodically by synchronization protocol, the residual TO drift within several chips.} and ${\Delta f_{i,{j_k}}} \in [-f_{\max},f_{\max}]$, $\forall {i \in {\left\{ {{{\cal P}_p}\backslash i_{j_k}^p} \right\}}}$, respectively. 

Obviously, the mean of ${{ \boldsymbol{\bar w}}_{i,{j_k}}}$ is $\boldsymbol{0}$ as all variables are independent distribution with zero mean. The variance of ${{\bf{F}}\left( {{\Delta f_{i,j_k}}} \right){{\bf{S}}_p}{\bf{U}}\left( {{\Delta \tau_{i,j_k}}} \right){{\bf{h}}_{i,j_k}}}$, $\forall {i \in {\left\{ {{{\cal P}_p}\backslash i_{j_k}^p} \right\}}}$, can be given by 
\begin{equation}
	\label{variance}
	\begin{split}
		&\text{var}\{{{\bf{F}}\left( {{\Delta f_{i,j_k}}} \right){{\bf{S}}_p}{\bf{U}}\left( {{\Delta \tau_{i,j_k}}} \right){{\bf{h}}_{i,j_k}}}\} \\
	=& \mathbb{E}\{ {{\bf{F}}\left( {{\Delta f_{i,j_k}}} \right){{\bf{S}}_p}{\bf{U}}\left( {{\Delta \tau_{i,j_k}}} \right){{\bf{h}}_{i,j_k}}}  ({{\bf{F}}\left( {{\Delta f_{i,j_k}}} \right){{\bf{S}}_p}{\bf{U}}\left( {{\Delta \tau_{i,j_k}}} \right){{\bf{h}}_{i,j_k}}} )^H\}  \\
		 =& \mathbb{E}\{ {{\bf{F}}\left( {{\Delta f_{i,j_k}}} \right){{\bf{S}}_p}{\bf{U}}\left( {{\Delta \tau_{i,j_k}}} \right){{\bf{h}}_{i,{j_k}}}}{{\bf{h}}^H_{i,{j_k}}} {\bf{U}}^H\left( {{\Delta \tau_{i,j_k}}} \right){{\bf{S}}_p^H} {\bf{F}}^H\left( {{\Delta f_{i,j_k}}}\right)\}.
	\end{split}	
\end{equation}

In the following, we aim to derive (\ref{variance}). To alleviate the pilot contamination, the distance among the slave APs that share the common pilot sequence should be as far as possible, resulting in severe path loss and lower received signal power. Furthermore, since the APs are deployed in relatively high positions, the APs can transmit signal without blockage issues. Based on these reasons, the pilot contamination from other clusters predominantly depends on the main path with path delay $\varsigma _{i,{j_k}}^0 = \frac{{{d_{i,{j_k}}}}}{c}$, $\forall {i \in {\left\{ {{{\cal P}_p}\backslash i_{j_k}^p} \right\}}}$, where $d_{i,{j_k}}$ is the distance between the $i$-th slave AP and the ${j_k}$-th master AP and $c$ is the speed of light. For this special case, ${\bf{U}}\left( {{\Delta \tau_{i,j_k}}} \right) \in {\mathbb R}^{(M-N+1) \times 1}$ is a vector and ${\bf h}_{i,j_k} \in {\mathbb C}^{1 \times 1}$ is $\alpha _{i,j_k}^0\sqrt {{\beta _{i,j_k}}} \psi^0_{i,j_k}$. Then, owing to the independent variables of ${\bf{F}}\left({{\Delta f_{i,j_k}}} \right)$ and ${\bf{U}}\left( {{\Delta \tau_{i,j_k}}} \right)$, we have
\begin{equation}
	\label{variance1}
	\begin{split}
		&\text{var}\{{{\bf{F}}\left( {{\Delta f_{i,j_k}}}  \right){{\bf{S}}_p}{\bf{U}}\left( {{\Delta \tau_{i,j_k}}}  \right){{\bf{h}}_{i,{j_k}}}}\} \\
		=& \alpha _{i,{j_k}}^0 {{\beta _{i,{j_k}}}}\mathbb{E}\Big\{ {{\bf{F}}\left( {{\Delta f_{i,j_k}}}  \right){{\bf{S}}_p} \mathbb{E}\{{\bf{U}}\left( {{\Delta \tau_{i,j_k}}}  \right)} {\bf{U}}^H \left( {{\Delta \tau_{i,j_k}}}  \right)\} {{\bf{S}}_p^H} {\bf{F}}^H\left( {{\Delta f_{i,j_k}}}  \right)\Big\}.
	\end{split}	
\end{equation}

Next, we derive the result of $\mathbb{E}\{ {\bf{U}}\left( {{\Delta \tau_{i,j_k}}} \right) {\bf{U}}^H\left( {{\Delta \tau_{i,j_k}}} \right)\}$, denoted as $\bf A$. Owing to the TO between the $i$-th slave AP and the ${j_k}$-th master AP, we have the following three cases. 

{\bf Case 1}: ${0 \le {{\Delta \tau_{i,j_k}}} < 1 - \frac{{\varsigma _{i,{j_k}}^0}}{{{T_s}}}}$, we define ${\bf \bar A}(0) = {\bf{U}}\left( {{\Delta \tau_{i,j_k}}} \right) {\bf{U}}^H\left( {{\Delta \tau_{i,j_k}}} \right)$ and have
\begin{equation}
	\label{case1}
	\begin{split}
		{\bf \bar A}(0) & = {\bf{U}}\left( {{\Delta \tau_{i,j_k}}} \right) {\bf{U}}^H\left( {{\Delta \tau_{i,j_k}}} \right)  
		= {[ {0,1 - x_{i,j_k}^0,x_{i,j_k}^0,0, \ldots ,0} ]^T}[ {0,1 - x_{i,j_k}^0,x_{i,j_k}^0,0, \ldots ,0} ],
	\end{split}
\end{equation}
where $x_{i,{j_k}}^0$ is $x_{i,{j_k}}^0 = {{\Delta \tau_{i,j_k}}} + \frac{{\varsigma _{i,{j_k}}^0}}{{{T_s}}}$. As can be seen from (\ref{case1}), there are only four non-zero elements, which are denoted as
\begin{equation}
	\label{case1_4}
	\begin{split}
		[{\bf \bar A}(0)]_{2,2} &= (1 - x_{i,{j_k}}^0)^2, 
		[{\bf \bar A}(0)]_{3,3} = (x_{i,{j_k}}^0)^2,
		[{\bf \bar A}(0)]_{2,3} = [{\bf \bar A}(0)]_{3,2} = (1-x_{i,{j_k}}^0)(x_{i,{j_k}}^0).
	\end{split}
\end{equation}

{\bf Case 2}: ${\upsilon - 1 - \frac{{\varsigma _{i,{j_k}}^0}}{{{T_s}}} \le {{\Delta \tau_{i,j_k}}} < \upsilon - \frac{{\varsigma _{i,{j_k}}^0}}{{{T_s}}}}$, $\forall \upsilon \in \{2,\cdots, {\eta}  \} $, we denote ${\bf{U}}\left( {{\Delta \tau_{i,j_k}}} \right) {\bf{U}}^H\left( {{\Delta \tau_{i,j_k}}} \right)$ as ${\bf \bar A}(\upsilon - 1) $ and have
\begin{equation}
	\label{case2}
	\begin{split}
		{\bf \bar A}(\upsilon - 1)& = {\bf{U}}\left( {{\Delta \tau_{i,j_l}}} \right) {\bf{U}}^H\left( {{\Delta \tau_{i,j_l}}} \right)  \\
		&= {[ {\underbrace {0, \cdots ,0}_\upsilon ,\upsilon  - x_{i,j_k}^0,x_{i,j_k}^0 + 1 - \upsilon , \ldots } ]^T}[ {\underbrace {0, \cdots ,0}_\upsilon ,\upsilon  - x_{i,j_k}^0,x_{i,j_k}^0 + 1 - \upsilon, \ldots } ].
	\end{split}
\end{equation}
Similar to the result given in (\ref{case1_4}), there are only four non-zero elements, which are denoted as
\begin{equation}
	\label{case2_4}
	\begin{split}
		[{\bf \bar A}(\upsilon - 1)]_{\upsilon+1,\upsilon+1} &= (\upsilon - x_{i,{j_k}}^0)^2, 
		[{\bf \bar A}(\upsilon - 1)]_{\upsilon+2,\upsilon+2} = (x_{i,{j_k}}^0+1-\upsilon)^2, \\
		[{\bf \bar A}(\upsilon - 1)]_{\upsilon+1,\upsilon+2} &= [{\bf \bar A}(\upsilon - 1)]_{\upsilon+2,\upsilon+1} = (\upsilon-x_{i,{j_k}}^0)(x_{i,{j_k}}^0+1-\upsilon).
	\end{split}
\end{equation}

{\bf Case 3}: ${\eta - \frac{{\varsigma _{i,{j_k}}^0}}{{{T_s}}} \le {{\Delta \tau_{i,j_k}}} < \eta }$, we define ${\bf \bar A}(\eta - \frac{{\varsigma _{i,{j_k}}^0}}{{{T_s}}}) = {\bf{U}}\left( {{\Delta \tau_{i,j_k}}} \right) {\bf{U}}^H\left( {{\Delta \tau_{i,j_k}}} \right)$ and have
\begin{equation}
	\label{case3}
	\begin{split}
	{\bf \bar A}(\eta - \frac{{\varsigma _{i,{j_k}}^0}}{{{T_s}}})=& {\bf{U}}\left({{\Delta \tau_{i,j_k}}} \right) {\bf{U}}^H\left( {{\Delta \tau_{i,j_k}}} \right)  \\
		= & {[ {\underbrace {0, \cdots }_{\eta+1} ,{\eta+1}  - x_{i,j_k}^0,x_{i,j_k}^0 - \eta , \ldots } ]^T}[ {\underbrace {0, \cdots}_{\eta+1} ,{\eta+1}  - x_{i,j_k}^0,x_{i,j_k}^0 - \eta , \ldots } ]
	\end{split}.
\end{equation}
The non-zero elements of ${\bf \bar A}(\eta - \frac{{\varsigma _{i,{j_k}}^0}}{{{T_s}}})$ are given by
\begin{equation}
	\label{case3_4}
	\begin{split}
		[{\bf \bar A}(\eta - \frac{{\varsigma _{i,{j_k}}^0}}{{{T_s}}})]_{\eta+2,\eta+2} &= (\eta +1- x_{i,{j_k}}^0)^2, 
		[{\bf \bar A}(\eta - \frac{{\varsigma _{i,{j_k}}^0}}{{{T_s}}})]_{\eta+3,\eta+3} = (x_{i,{j_k}}^0 - \eta)^2, \\
		[{\bf \bar A}(\eta - \frac{{\varsigma _{i,{j_k}}^0}}{{{T_s}}})]_{\eta+2,\eta+3} &= [{\bf \bar A}(\eta - \frac{{\varsigma _{i,{j_k}}^0}}{{{T_s}}})]_{\eta+3,\eta+2} = (\eta + 1-x_{i,{j_k}}^0)(x_{i,{j_k}}^0-\eta).
	\end{split}
\end{equation}

By using the integration and the above results, $\mathbb{E}\{ {\bf{U}}\left( {{\Delta \tau_{i,j_k}}} \right) {\bf{U}}^H\left( {{\Delta \tau_{i,j_k}}} \right)\}$ is given by
\begin{equation}
	\label{UUH}
	\begin{split}
	{\bf A} = &\mathbb{E}\{ {\bf{U}}\left( {{\Delta \tau_{i,j_k}}} \right) {\bf{U}}^H\left( {{\Delta \tau_{i,j_k}}} \right)\}  
	  = \mathbb{E}\{ {\bf \bar A}(0)+ \sum_{v=2}^{\eta}{\bf \bar A}(\upsilon-1) + {\bf \bar A}(\eta- \frac{{\varsigma _{i,{j_k}}^0}}{{{T_s}}}) \}.
	\end{split}
\end{equation}
Based on these discussions, if $\eta > 1$, the elements of ${\bf A}$ are given by
\begin{equation}
	\setlength\abovedisplayskip{5pt}
	\setlength\belowdisplayskip{5pt}
	\label{Aelement}
	\begin{split}
		[{\bf A}]_{2,3} =[{\bf A}]_{3,2} = {\frac{{1 - 3{{\left( {\frac{{\varsigma _{i,{j_k}}^0}}{{{T_s}}}} \right)}^2} + 2{{\left( {\frac{{\varsigma _{i,{j_k}}^0}}{{{T_s}}}} \right)}^3}}}{6\eta}},
	\end{split}
\end{equation}
\begin{equation}
	\setlength\abovedisplayskip{5pt}
	\setlength\belowdisplayskip{5pt}
	\label{A33}
	\begin{split}
		[{\bf A}]_{2,2} &= {\frac{(1-{\frac{{\varsigma _{i,{j_k}}^0}}{{{T_s}}}})^3}{3\eta}},[{\bf A}]_{3,3} = {\frac{{2 - {{\left( {\frac{{\varsigma _{i,{j_k}}^0}}{{{T_s}}}} \right)}^3}}}{3\eta}},\\
		[{\bf A}]_{\upsilon+1,\upsilon+2} &= [{\bf A}]_{\upsilon+2,\upsilon+1} = \frac{1}{6\eta}, \forall \upsilon \in \{2,\cdots,\eta\},
	\end{split}
\end{equation}
\begin{equation}
	\setlength\abovedisplayskip{5pt}
	\setlength\belowdisplayskip{5pt}
	\begin{split}	
		[{\bf A}]_{\upsilon+2,\upsilon+2} = \frac{2}{3\eta}, \forall \upsilon \in \{2,\cdots,\eta-1\}, 
	\end{split}
\end{equation}
\begin{equation}
	\setlength\abovedisplayskip{5pt}
	\setlength\belowdisplayskip{5pt}
	\begin{split}	
		[{\bf A}]_{\eta+2,\eta+3} &= [{\bf A}]_{\eta+3,\eta+2} = {\frac{{3{{\left( {\frac{{\varsigma _{i,{j_k}}^0}}{{{T_s}}}} \right)}^2} - 2{{\left( {\frac{{\varsigma _{i,{j_k}}^0}}{{{T_s}}}} \right)}^3}}}{6\eta}},
	\end{split}
\end{equation}
\begin{equation}
	\setlength\abovedisplayskip{5pt}
	\setlength\belowdisplayskip{5pt}
	\begin{split}	
		[{\bf A}]_{\eta+2,\eta+2} \!&=\! \frac{2 \!-\! (1\!-\!{\frac{{\varsigma _{i,{j_k}}^0}}{{{T_s}}}})^3}{3\eta}, [{\bf A}]_{\eta+3,\eta+3}\! =\! {\frac{{{{\left( {\frac{{\varsigma _{i,{j_k}}^0}}{{{T_s}}}} \right)}^3}}}{3\eta}}.
	\end{split}
\end{equation}

For the special case of $\eta = 1$, the $[3,3]$-th element of ${\bf A}$ is different from the result given in (\ref{A33}), which is given by 
\begin{equation}
	\label{A331}
	[{\bf A}]_{3,3} = {\frac{{1 - 3{{\left( {\frac{{\varsigma _{i,{j_k}}^0}}{{{T_s}}}} \right)}^2} + 3{\frac{{\varsigma _{i,{j_k}}^0}}{{{T_s}}}}}}{3}}.
\end{equation}

Then, since the $[n,m]$-th element of ${\bf{F}}\left( {{\Delta f_{i,j_k}}}  \right){{\bf{S}}_p}$ is $e^{j2\pi(n-1){\Delta f_{i,{j_k}}}{T_s}} {s}_p^{n-m}$, we have
\begin{equation}
	\label{Z}
	\begin{split}
			&[{{\bf{F}}\left( {{\Delta f_{i,j_k}}}  \right){{\bf{S}}_p} \mathbb{E}\{{\bf{U}}\left( {{\Delta \tau _{i,{j_k}}}} \right)} {\bf{U}}^H \left( {{\Delta \tau_{i,j_k}}}  \right)\}]_{n,m} = \sum_{k'=1}^{M-N+1} {[{\bf{F}}\left( {{\Delta f_{i,j_k}}}  \right){{\bf{S}}_p}]_{n,k'} [{\bf A}]_{k',m}} \\
			= & \sum_{k'=m-1}^{m+1}{[{\bf A}]_{k',m}e^{j2\pi({n-1}) {\Delta f_{i,{j_k}}}{T_s}}{ s}_p^{n-k'}},
	\end{split}
\end{equation}
where $s^{n-k'}_p$ is $0$ when $n-k' < 0$ and $n-k' > N-1$.

By substituting the result of (\ref{Z}) into (\ref{variance1}), we have
\begin{equation}
	\begin{split}
		& [{{\bf{F}}\left( {{\Delta f_{i,j_k}}}  \right){{\bf{S}}_p} \mathbb{E}\{{\bf{U}}\left( {{\Delta \tau_{i,j_k}}}  \right)} {\bf{U}}^H \left( {{\Delta \tau_{i,j_k}}}  \right)\} {{\bf{S}}_p^H} {\bf{F}}^H\left( {{\Delta f_{i,j_k}}}  \right)]_{n,m} \\
		= & \sum\limits_{k' = 1}^{M-N+1} {[{{\bf{F}}\left( {{\Delta f_{i,j_k}}}  \right){{\bf{S}}_p} \mathbb{E}\{{\bf{U}}\left( {{\Delta \tau_{i,j_k}}}  \right)} {\bf{U}}^H \left( {{\Delta \tau_{i,j_k}}}  \right)\}]_{n,k'} [{{\bf{S}}_p^H} {\bf{F}}^H\left( {{\Delta f_{i,j_k}}}  \right)]_{k',m}} \\
		= & \sum\limits_{k' = 1}^{M-N+1} { \sum_{k_1=k'-1}^{k'+1}{[{\bf A}]_{k_1,k'}e^{j2\pi({n-1}) {\Delta f_{i,{j_k}}}{T_s}}{ s}_p^{n-k_1}} e^{-j2\pi(m-1){\Delta f_{i,{j_k}}}{T_s}} {s}_p^{m-k'}} \\
		= & \sum\limits_{k' = 2}^{M-N+1} { \sum_{k_1=k'-1}^{k'+1}{[{\bf A}]_{k_1,k'}e^{j2\pi({n-m}) {\Delta f_{i,{j_k}}}{T_s}}{ s}_p^{n-k_1}} {s}_p^{m-k'}}.
	\end{split}
\end{equation}
Then, based on the integration of CFO $\Delta f_{i,{j_k}}$, we have 
\begin{equation}
	\begin{split}
		\int_{-f_{\max}}^{f_{\max}} { \frac{e^{j2\pi(n-m) \Delta f_{i,{j_k}}T_s}}{2 f_{\max}}  d_{\Delta f_{i,{j_k}}}  = \left\{ {\begin{array}{*{20}{c}}
					{1,}&{m = n}\\
					{0,}&{m \ne n}.
			\end{array}} \right.} 
	\end{split}
\end{equation}  
As a result, $\text{var}\{{{\bf{F}}\left( {{\Delta  f_{i,{j_k}}}} \right){{\bf{S}}_p}{\bf{U}}\left( {{\Delta \tau_{i,{j_k}}}} \right){{\bf{h}}_{i,{j_k}}}}\}$ is a diagonal matrix, denoted as ${\bf D}_{i,{j_k}}$, whose $[n,n]$-th element can be expressed as
\begin{equation}
	\label{FSU}
	\begin{split}
	[{{\bf D}_{i,{j_k}}}]_{n,n} & \buildrel \Delta \over =	[\text{var}\{{{\bf{F}}\left( {{\Delta f_{i,j_k}}}  \right){{\bf{S}}_p}{\bf{U}}\left( {{\Delta \tau_{i,j_k}}}  \right){{\bf{h}}_{i,{j_k}}}}\}]_{n,n} \\
		& = \alpha _{i,{j_k}}^0 {{\beta _{i,{j_k}}}} \sum\limits_{k' = 2}^{M-N+1} { \sum_{k_1=k'-1}^{k'+1}{[{\bf A}]_{k_1,k'}{ s}_p^{n-k_1}} {s}_p^{n-k'}}. 
	\end{split}
\end{equation}

By substituting (\ref{FSU}) into (\ref{variance1}), we obtain $\boldsymbol{\Xi}$, which can be written as
\begin{equation}
	\label{Xi}
	\begin{split}
		\boldsymbol{\Xi} = \sum\limits_{i \in {\left\{ {{{\cal P}_p}\backslash i_{j_k}^p} \right\}}} {{{\bf{D}}_{i,{j_k}}}}  + {\sigma ^2}{\bf{I}}_M.
	\end{split}
\end{equation}

In the following, we derive the partial derivatives of $\boldsymbol{\theta}$. Define ${\bf M} = {\text{diag}}(0,1,\cdots, M)$ and let ${\bf e}_{l}^{i^p_{j_k},{j_k}}$ denote the $l$-th column of unit matrix ${\bf I}_{L_{i^p_{j_k},{j_k}}}$. Denote ${\bf{U}}'\left( {{\Delta \tau_{i^p_{j_k},{j_k}}}} \right)$ as the partial derivation with respect to ${{\Delta \tau_{i^p_{j_k},{j_k}}}}$, whose $l$-th column has two non-zeros elements equal to $-1$ at component $\left\lceil { 1+ \frac{{{\Delta \tau _{i^p_{j_k},{j_k}}}}}{{{T_s}}} + \frac{{\varsigma _{i^p_{j_k},{j_k}}^l}}{{{T_s}}}} \right\rceil $ and equal to $1$ at component $1 + \left\lceil { 1+\frac{{{\Delta \tau _{i^p_{j_k},{j_k}}}}}{{{T_s}}} + \frac{{\varsigma _{i^p_{j_k},{j_k}}^l}}{{{T_s}}}} \right\rceil $. Based on these definitions, we have
\begin{equation}
	\setlength\abovedisplayskip{5pt}
	\setlength\belowdisplayskip{5pt}
	\label{f}
	\frac{{\partial {\bf{m}}\left( \boldsymbol{\theta}  \right)}}{{\partial \theta_1}} =  {\bf{F}}\left( {\Delta {f_{i_{j_k}^p,{j_k}}}} \right){{\bf{S}}_p}{\bf{U}}'\left( {\Delta {\tau _{i_{j_k}^p,{j_k}}}} \right){{\bf{h}}_{i_{j_k}^p,{j_k}}},
\end{equation}
\begin{equation}
	\setlength\abovedisplayskip{5pt}
	\setlength\belowdisplayskip{5pt}
	\label{tau}
	\frac{{\partial {\bf{m}}\left( \boldsymbol{\theta}  \right)}}{{\partial \theta_2}} \!= \! j2\pi\! {\bf{MF}}\!\left( {\Delta {f_{i_{j_k}^p,{j_k}}}} \right)\!{{\bf{S}}_p}{\bf{U}}\!\left( {\Delta {\tau _{i_{j_k}^p,{j_k}}}} \right)\!{{\bf{h}}_{i_{j_k}^p,{j_k}}},
\end{equation}
\begin{equation}
	\setlength\abovedisplayskip{5pt}
	\setlength\belowdisplayskip{5pt}
	\label{real}
	\begin{split}
		&\frac{{\partial {\bf{m}}\left( \boldsymbol{\theta}  \right)}}{{\partial \theta_{2l+3}}}\! = \! {\bf{F}}\!\left( {\Delta {f_{i_{j_k}^p,{j_k}}}} \right)\!{{\bf{S}}_p}{\bf{U}}\!\left( {\Delta {\tau _{i_{j_k}^p,{j_k}}}} \right)\!{{\bf{h}}_{i_{j_k}^p,j_k}}{\bf e}_{l}^{i_{j_k}^p,{j_k}},
	\end{split}
\end{equation}
\begin{equation}
	\setlength\abovedisplayskip{5pt}
	\setlength\belowdisplayskip{5pt}
	\label{imga}
	\begin{split}
		&\frac{{\partial {\bf{m}}\left( \boldsymbol{\theta}  \right)}}{{\partial \theta_{2l+4}}} \!=\! j {\bf{F}}\!\left( {\Delta {f_{i_{j_k}^p,{j_k}}}} \right)\!{{\bf{S}}_p}{\bf{U}}\!\left( {\Delta {\tau _{i_{j_k}^p,{j_k}}}} \right)\!{{\bf{h}}_{i_{j_k}^p,{j_k}}}{\bf e}_{l}^{i_{j_k}^p,{j_k}}, \\ & \qquad \qquad \forall l \in \{0,1,\cdots, L_{i_{j_k}^p,{j_k}} -1\}.
	\end{split}
\end{equation}
Finally, by defining the $M \times 2(L_{i_{j_k}^p,{j_k}} + 1)$ matrix $ \boldsymbol{\Omega}  = [\frac{{\partial {\bf{m}}\left( \boldsymbol{\theta}  \right)}}{{\partial \theta_1}},\frac{{\partial {\bf{m}}\left( \boldsymbol{\theta}  \right)}}{{\partial \theta_2}},\cdots,\frac{{\partial {\bf{m}}\left( \boldsymbol{\theta}  \right)}}{{\partial \theta_{2L_{i_{j_k}^p,{j_k}}+2}}}]$, the Fisher matrix can be written as  
\begin{equation}
	\setlength\abovedisplayskip{5pt}
	\setlength\belowdisplayskip{5pt}
	\label{finfisher}
	{\bf J}(\boldsymbol{\theta}) = 2 \text{Re}\{{\boldsymbol \Omega}^H \boldsymbol{\Xi} {\boldsymbol \Omega} \}.
\end{equation}
Finally, the CRBs of CFO and TO are obtained by the first two diagonal elements of ${\bf J}^{-1}(\boldsymbol{\theta})$. These CRBs can be regarded as the theoretical lower bound of mean square errors (MSEs) between the real parameters and estimated results. With the given result in (\ref{finfisher}), the synchronization performance depends on the matrix $\boldsymbol{\Xi}$. For the low SNR regime, both the pilot contamination and noise power have significantly impacts on synchronization errors. For the high SNR regime, the power $\boldsymbol{\Xi}$ mainly comes from the pilot contamination, and thus it is worth investigating the pilot allocation strategy to improve the synchronization performance and cooperative ISAC performance.
%

\subsection{Maximum Likelihood Estimation}
In this subsection, we jointly estimate CFO and TO between the $i_{j_k}^p$ slave AP and the ${j_k}$-th master AP, assuming that the noise power $\sigma^2$ and the path delays $\varsigma_{i_{j_k}^p,{j_k}}^l$, $\forall l \in \{0,1,2,...,L_{i_{j_k}^p,{j_k}}-1\}$ are known \footnote{Due to the fact that the $i_{j_k}^p$-th slave AP and the ${j_k}$-th master AP are in the same cluster, the distances from the $i_{j_k}^p$-th slave AP to the ${j_k}$-th master AP is closer than that of the slave APs in other clusters and the variation of multi-path delay is relatively slow compared to inter-cluster ones. Furthermore, since the synchronization is performed periodically in a short time interval and all APs are deployed in high locations with relatively stable propagation paths, these factors lead to the known path delays.}. Then, the ML estimation can be obtained by minimizing the square distance, which is given by
\begin{equation}
	\label{LF}
	\left\| {{{\bf{y}}_{j_k}} - {\bf{F}}\left( {\Delta {f_{i_{j_k}^p,{j_k}}}} \right){{\bf{S}}_p}{\bf{U}}\left( {\Delta {\tau _{i_{j_k}^p,{j_k}}}} \right){{\bf{h}}_{i_{j_k}^p,{j_k}}}} \right\|^2.
\end{equation}

For given $\frac{{\Delta {\tau _{i_{j_k}^p,{j_k}}}}}{{{T_s}}}$ and $\Delta {f_{i_{j_k}^p,{j_k}}}{T_s}$, the estimation of ${\bf h}_{i_{j_k}^p,{j_k}}$ can be obtained by
\begin{equation}
	\label{estimateh}
	\begin{split}
	{\bf \hat h}_{i_{j_k}^p,{j_k}} &= \Big({\bf{U}}^H\left( {\Delta {\tau _{i_{j_k}^p,{j_k}}}} \right){{\bf{S}}_p^H}{\bf{F}}^H\left( {\Delta {f_{i_{j_k}^p,{j_k}}}} \right) {\bf{F}}\left( {\Delta {f_{i_{j_k}^p,{j_k}}}} \right){{\bf{S}}_p}{\bf{U}}\left( {\Delta {\tau _{i_{j_k}^p,{j_k}}}} \right)\Big)^{-1}\\
	& \quad \times {\bf{U}}^H\left( {\Delta {\tau _{i_{j_k}^p,{j_k}}}} \right){{\bf{S}}_p^H}{\bf{F}}^H\left( {\Delta {f_{i_{j_k}^p,{j_k}}}} \right) {\bf y}_{j_k}.
	\end{split}
\end{equation}
Then, by substituting the result of (\ref{estimateh}) into (\ref{LF}), we can estimate $\Delta f_{i_{j_k}^p,{j_k}}T_s$ and  $\frac{\Delta \tau_{i_{j_k}^p,{j_k}}}{T_s}$ by maximizing the following quadratic form 
\begin{equation}
	\label{estimateft}
	\begin{split}
		& {\bf y}_{j_k} {\bf{F}}\left( {\Delta {f_{i_{j_k}^p,{j_k}}}} \right){{\bf{S}}_p}{\bf{U}}\left( {\Delta {\tau _{i_{j_k}^p,{j_k}}}} \right) \Big({\bf{U}}^H\left( {\Delta {\tau _{i_{j_k}^p,{j_k}}}} \right){{\bf{S}}_p^H}{\bf{F}}^H\left( {\Delta {f_{i_{j_k}^p,{j_k}}}} \right) {\bf{F}}\left( {\Delta {f_{i_{j_k}^p,{j_k}}}} \right){{\bf{S}}_p}{\bf{U}}\left( {\Delta {\tau _{i_{j_k}^p,{j_k}}}} \right)\Big)^{-1} \\
		 \times & {\bf{U}}^H\left( {\Delta {\tau _{i_{j_k}^p,{j_k}}}} \right){{\bf{S}}_p^H}{\bf{F}}^H\left( {\Delta {f_{i_{j_k}^p,{j_k}}}} \right) {\bf y}_{j_k}.
	\end{split}
\end{equation}
This can be regarded as maximizing the energy of the projection ${\bf y}_{j_k}$ onto the column space of the signal
matrix ${\bf{F}}\left( {\Delta {f_{i_{j_k}^p,{j_k}}}} \right){{\bf{S}}_p}{\bf{U}}\left( {\Delta {\tau _{i_{j_k}^p,{j_k}}}} \right)$. As a result, the ML estimator is obtained by searching over a two-dimensional sufficiently fine grid of points with respect to the variables $\Delta f_{i_{j_k}^p,{j_k}}T_s$ and  $\frac{\Delta \tau_{i_{j_k}^p,{j_k}}}{T_s}$.	
\section{Synchronization based on Pilot Assignment}
Based on the result given in (\ref{finfisher}), we devise a synchronization scheme based on pilot sharing to minimize the sum of CRBs while simultaneously satisfying the requirement for cooperative ISAC with limited overhead.  

\subsection{Problem Formulation}
In this subsection, we jointly optimize the cluster classification, pilot overhead, and pilot sharing scheme to minimize the synchronization errors. By denoting the synchronization requirement of the $i$-th slave AP and the $j_k$-th master AP as ${\rm CRB}^{\rm Req}_{i,j_k}$, $j_k \in {\mathcal C}_k$, $\forall i \in \{{\mathcal C}_k \backslash j_k\}$, $k = \{1,\cdots, K_{C}\}$, the minimization of sum CRBs can be formulated as (\ref{optproblem}), which is given by
\begin{subequations}
	\label{optproblem}
	\begin{align}
		 \mathop {\min }\limits_{K_C,{\mathcal{C}_k, k \in \left\{1,\cdots, K_{C}\right\}}, {\tau}, {{{\cal P}_p},p \in \left\{ {1, \cdots ,{\tau}} \right\}}}& 	\sum\limits_{k = 1 }^{K_C} {\sum\limits_{i \in \{{\cal C}_k \backslash j_k\}} {{{\rm{CRB}}_{i,j_k}(\Delta f_{i,j_k},\Delta \tau_{i,j_k})}} } \notag \\
		  \text{s. t.} & \qquad {\rm CRB}_{i,j_k}(\Delta f_{i,j_k},\Delta \tau_{i,j_k}) \leq {\rm CRB}_{i,j_k}^{\rm Req}, \notag\\
		& \qquad \qquad j_k \in {\mathcal C}_k,\forall i \in \{{\mathcal C}_k \backslash j_k\}, k \in \{1,\cdots, K_{C}\}, \label{constrain1} \\ 
		& \qquad \qquad 2K_C + \tau \leq L_S. \label{constrain2}
	\end{align}
\end{subequations}
In (\ref{optproblem}), ${{{\rm{CRB}}_{i,j_k}(\Delta f_{i,j_k},\Delta \tau_{i,j_k})}}$ is the theoretical lower bound for estimation errors of CFO and TO between the $i$-th slave AP and the $j_k$-th master AP, $\tau$ is the pilot overhead, and $L_S$ is the total overhead for synchronization.
Constraint (\ref{constrain1}) means that each AP should satisfy the minimal requirements for synchronization and cooperative ISAC, and constraint (\ref{constrain2}) represents that the total overhead, including the pilot overhead and information exchange, should be less than $L_S$.

As can be seen from (\ref{optproblem}), the synchronization performance will be better if there are more master APs and only one slave AP. However, it is unrealistic to deploy more master APs and only one slave AP in the cell-free mMIMO systems, which leads to significant overhead for exchanging information and impractical implementation. Furthermore, considering the limited channel resources and high requirements of synchronization, it is challenging to strike a balance between the synchronization performance and overhead, i.e., information exchange and pilot overhead. To address this issue, we aim to devise a synchronization scheme with low overhead while simultaneously satisfying the requirements for cooperative ISAC. 

\begin{figure}[t]
	\centering
	\includegraphics[width=4.5 in]{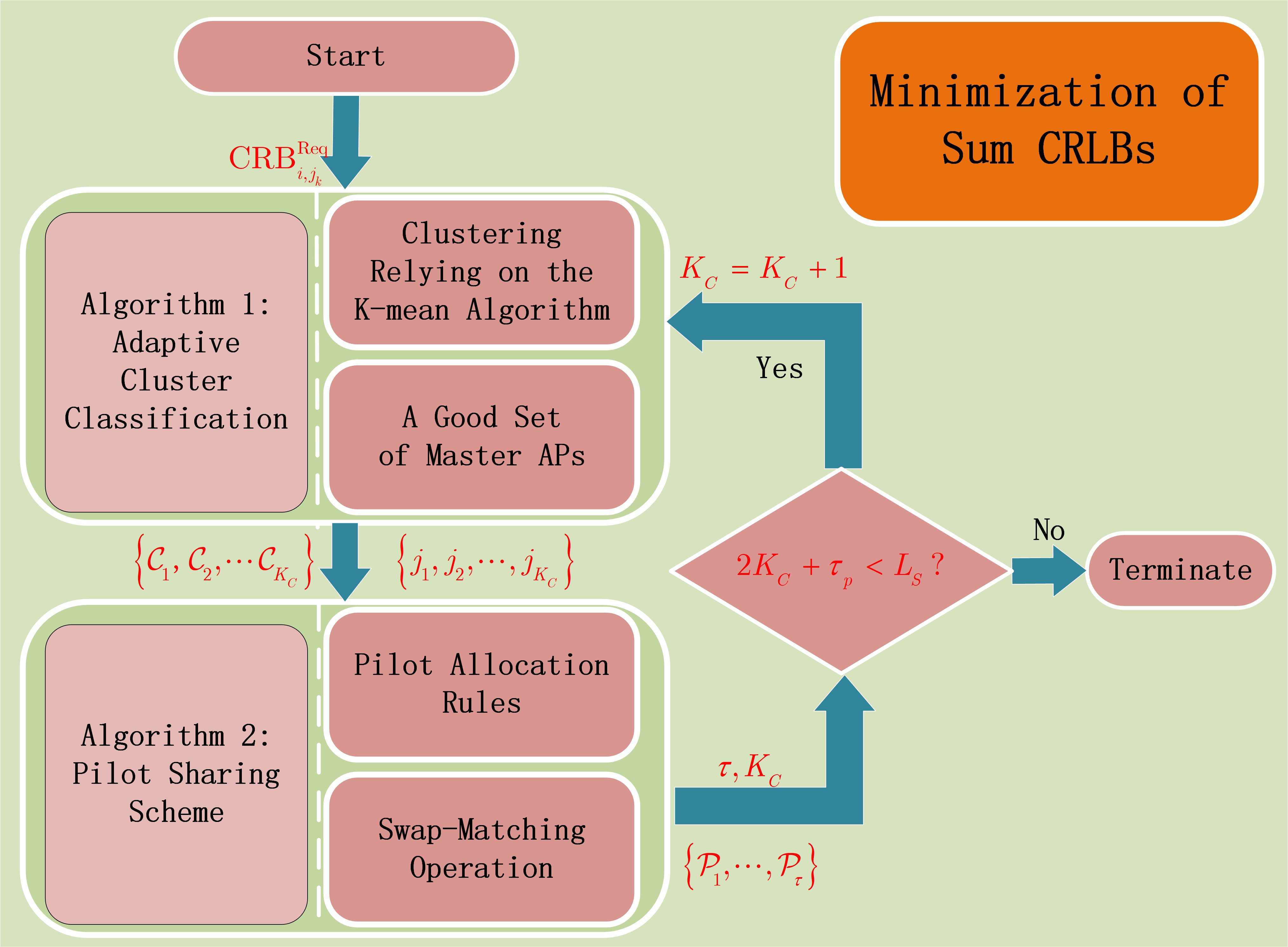}
	\caption{Illustration for solving the sum of CRBs.}
	\label{structure}
\end{figure}

To address this NP-hard problem, we first need to classify the clusters according to APs' synchronization requirements, and then allocate the pilot sequences to all slave APs. Therefore, Problem (\ref{optproblem}) can be simplified into two sub-problems, including adaptive cluster classification problem and pilot sharing scheme, which are depicted in Fig. \ref{structure}.

\subsection{Adaptive Cluster Classification}
In this subsection, we find a good way to classify the clusters with a good set of master APs while considering the overhead of information exchange. By ignoring the pilot allocation strategy, Problem (\ref{optproblem}) can be simplified into
\begin{subequations}
	\label{subproblem1}
	\begin{align}
		\mathop {\min }\limits_{K_C, {\mathcal{C}_k, k \in \left\{1,\cdots, K_{C}\right\}}}& 	\sum\limits_{k =1}^{K_C} {\sum\limits_{i \in \{{\cal C}_k \backslash j_k\}} {{{\rm{CRB}}_{i,j_k}(\Delta f_{i,j_k},\Delta \tau_{i,j_k})}} } \notag \\
		\text{s. t.} & \qquad {\rm CRB}_{i,j_k}(\Delta f_{i,j_k},\Delta \tau_{i,j_k}) \leq {\rm CRB}_{i,j_k}^{\rm Req}, \notag\\
		& \qquad \qquad j_k \in {\mathcal C}_k,\forall i \in \{{\mathcal C}_k \backslash j_k\}, l \in \{1,\cdots, K_{C}\}, \label{subconstrain11} \\
		& \qquad \qquad 2K_C < L_S. \label{subconstrain12}
	\end{align}
\end{subequations}
Obviously, minimizing the sum of CRBs can be obtained by simply having one slave AP with the largest path gain, but this violates the constraint in (\ref{subconstrain12}) owing to the heavy burden on information exchange. Conversely, deploying an insufficient number of clusters may increase the average distance between master AP and slave APs in each cluster, which leads to a significant decrease in synchronization performance. Furthermore, even though the clusters are classified, it is still worth exploring which AP can be chosen as the master AP to minimize the CRB. To address these issues, an adaptive cluster classification based on the synchronization (or ISAC) requirements is proposed, and a criterion that finds a good set of master APs is to minimize the sum of CRBs.

In the following, our first step is to investigate how to find a reasonable number of clusters according to the stringent synchronization performance. Since estimation errors decrease with the signal-to-noise ratio (SNR) (dB) \cite{rogalin2014scalable}, we can similarly transform the requirement of synchronization to the requirement of signal-to-interference-plus-noise ratio (SINR) (dB), by using the following theorem.
\begin{theorem}
	\label{CRB2SINR}
	By using the binary phase shift keying (BPSK)  modulation and random pilot sequence, we have 
	\begin{equation}
		\label{averagepc}
		\begin{split}
			\mathbb{E}\{\boldsymbol{\Xi}\} = (\sum\limits_{i \in \left\{ {{{\cal P}_p}\backslash i_{{j_k}}^p} \right\}} {\alpha _{i,{j_k}}^0{\beta _{i,{j_k}}}\sum\limits_{k' = 2}^{M - N + 1} { {{{[{\bf{A}}]}_{{k'},k'}}} } }) {\bf I}_{M} + {\sigma ^2}{\bf I}_{M}.
		\end{split}
	\end{equation} 
	
	{\emph{Proof}}: Owing to BSPK modulation and the fact of ${\bf A} = {\bf A}^H$, we have 
	\begin{equation}
		\label{averagexi}
		\begin{split}
			\mathbb{E}\{\sum\limits_{k' = 2}^{M - N + 1}{\sum\limits_{{k_1} = k' - 1}^{k' + 1} {{{[{\bf{A}}]}_{{k_1},k'}}s_p^{m' - {k_1}}} s_p^{m' - k'}}\} 
			= \sum_{k' = 2}^{M-N+1} {{{[{\bf{A}}]}_{{k'},k'}}}.
		\end{split}
	\end{equation}
	Therefore, we complete this proof.
\end{theorem}

Therefore, the Fish matrix is related to the average power of pilot contamination and noise. As a result, based on the above discussion, the CRB can be equivalently transformed into SINR, which can be expressed as	
\begin{equation}
	\label{MSE2SINR}
	\begin{split}
		{\rm CRB}_{i,j_k}(\Delta f_{i,j_k},\Delta \tau_{i,j_k}) \leq {\rm CRB}_{i,j_k}^{\rm Req}
		\Rightarrow {\rm SINR}_{i,j_k}  \ge  {\rm SINR}_{i,j_k}^{\rm Req}.
	\end{split}
\end{equation}
Then, based on the required SINR, we derive the size of each cluster, which can be used to determine the number of clusters. Specifically, the maximum intra-cluster distance can be found by relying on the relationship between the distance and path loss. For the ideal case, i.e., no multiple paths and no pilot contamination, the expectation of ideal SINR can be written as 
\begin{equation}
	\label{idealSINR}
	{\rm SINR}^{\rm ideal} =10\log_{10}\Big(\frac{{\rm Tr}\{{\bf D}^{\rm ideal}\}}{ {M\sigma^2 }}\Big),
\end{equation}
where ${\bf D}^{\rm ideal}$ is $\text{var}\Big\{{{\bf{F}}\left( {{\Delta f_{i^p_{j_k},j_k}}}  \right){{\bf{S}}_p}{\bf{U}}\left( {{\Delta \tau_{i^p_{j_k},j_k}}} \right){{\bf{h}}_{i^p_{j_k},{j_k}}}}\Big\}$, whose proof is omitted owing to the same derivation of (\ref{FSU}). Then, the maximum intra-cluster distance is determined by the minimum ${{\rm MSE}}^{\rm Req}_{i,j_k}$, $\forall i \in \{{\mathcal C}_k \backslash j_k\}$, $\forall k \in \{1,\cdots,K_C\}$. For ease of expression, we denote the ${\rm SINR}^{\min}$ as the equivalent condition of the minimum $ {{\rm MSE}}^{\rm Req}_{i,j_k}$. By assuming that $\alpha_{i^p_{j_k},{j_k}}^0$ is 1 and substituting (\ref{idealSINR}) into (\ref{MSE2SINR}), we have 
\begin{equation}
	\label{mindis}
	\begin{split}
		& {\rm SINR}^{\rm ideal} \ge  {\rm SINR}^{\min} \Rightarrow \beta^{\min} \ge \frac{M\sigma^2 10^{\frac{{\rm SINR}^{\min}}{10}}}{ {\rm Tr} \Big\{ {\mathbb E}\big\{ {\bf F}(\Delta f_{i^p_{j_k},j_k}) {\bf S}_p {\bf A}{\bf S}_p^H {\bf F}^H(\Delta f_{i^p_{j_k},j_k})\big\}\Big\}}.
	\end{split}
\end{equation}
By using the three-slope channel model given in \cite{Ngo2017}, the maximum intra-cluster distance can be readily obtained by
\begin{equation}
	\label{maxdis}
	\begin{split}
		{\rm dis}^{\max} = {\rm{1}}{{\rm{0}}^{ - \frac{2}{7}\log \frac{{M\sigma^210^{\frac{{\rm SINR}^{\min}}{10}}}}{ {\rm Tr} \Big\{ {\mathbb E}\{ {\bf F}(\Delta f_{i^p_{j_k},j_k}) {\bf S}_p {\bf A}{\bf S}_p^H {\bf F}^H(\Delta f_{i^p_{j_k},j_k})\}\Big\}} - \frac{L^{\rm loss}}{{35}}}},
	\end{split}
\end{equation}
where $L^{\rm loss}$ is the constant factor. Although we cannot directly determine the number of clusters, we can use this maximum intra-cluster distance to determine whether the clusters satisfy the synchronization requirements.

Based on the above discussions, for a fixed number of cluster $K_C$, the minimum sum of CRBs can be obtained by maximizing the sum of SINRs, which can be expressed as 
\begin{subequations}
	\label{subproblem11}
	\begin{align}
		\mathop {\max }\limits_{{\mathcal{C}_k, k \in \left\{1,\cdots, K_{C}\right\}}}& 	\sum\limits_{k =1}^{K_C} {\sum\limits_{i \in \{{\cal C}_k \backslash j_k\}} {{{\rm{SINR}}_{i,j_k}}} } \notag \\
		\text{s. t.} & \qquad {\rm SINR}_{i,j_k} \ge  {\rm SINR}_{i,j_k}^{\rm Req}, j_k \in {\mathcal C}_k,\forall i \in \{{\mathcal C}_k \backslash j_k\}, k \in \{1,\cdots, K_{C}\}, \label{subconstrain111} \\
		& \qquad \qquad 2K_C < L_S. \label{subconstrain121}
	\end{align}
\end{subequations}
Although it is impossible to derive a closed-form expression of ${\rm SINR}_{i,j_k}$, the result of the pilot contamination reveals that the power primarily depends on distance-based large-scale fading factors. Consequently, the maximum sum of SINRs can be obtained by shortening the distance between the slave AP and master AP. Then, the cluster can be classified by minimizing the sum of intra-cluster distances while ensuring that the AP-to-centroid distance in the cluster is no larger than the maximum intra-cluster distance ${\rm dis}^{\max}$, which can be written as
	\begin{subequations}
		\label{classification}
		\begin{align}
			\min \limits_{K_C,{\cal C}_1, \cdots, {\cal C}_{K_C}}& \sum_{k=1}^{K_C}{\sum_{i \in {\cal C}_k}} |(x_k,y_k) - {\rm Loc}_i|, \notag \\
			\text{s. t.} & \qquad |(x_k,y_k) - {\rm Loc}_i| \leq {\rm dis}^{\max},\forall i \in {\mathcal C}_k, k \in \{1,\cdots, K_{C}\}, \label{classification1} 
		\end{align}	
	\end{subequations}
	where $(x_k,y_k)$ means the optimal 2-dimensional location of the $k$-th cluster's centroid that has the minimal sum of distances to all APs in the $k$-th cluster, and ${\rm Loc}_i$ is the 2-dimensional location of the $i$-th AP.  However, for the given number of clusters $K_C$, it is still challenging to obtain the optimal solution of Problem (\ref{classification}) since it is an the NP-hard problem. To address this issue, we classify the clusters by using the K-means algorithm, whose details are given in \cite{kanungo2002efficient}. By combining the K-means algorithm with the maximum intra-cluster distance, we can gradually classify the clusters by gradually increasing the number of clusters $K_C$ until the maximum intra-cluster distances of all clusters are no more than the value given in (\ref{maxdis}).


Based on the given clusters, a good set of master APs should be found to minimize the sum of CRBs. Similar to the clustering approach, the master AP and the centroid of the cluster should have the same characteristics, thereby maximizing the sum of path gains in the cluster and improving the synchronization performance. Mathematically, a criteria that finds a good set of master APs can be written as  
\begin{equation}
	\label{masterap}
	{j_k} = \mathop {\arg \min }\limits_{{j_k} \in {{\cal C}_k}} \left\{ {\sum\limits_{i \in \left\{ {{{\cal C}_k}\backslash {j_k}} \right\}} {\left\| {{{\rm{Loc}}_{{j_k}}} - {{\rm{Loc}}_i}} \right\|} } \right\},\forall k \in \left\{ {1, \cdots ,{K_C}} \right\}.
\end{equation} 
By using this criterion of (\ref{masterap}), the master AP has the minimum sum distances to all slave APs, which leads to larger path gain and better synchronization performance.

Based on the above discussions, we propose an adaptive cluster classification based on the K-means algorithm, which is detailed in Algorithm \ref{kmeans}. Initially, we divide all APs in this area into $K_C$ clusters and find the maximum intra-cluster distance between the master AP and the slave AP of each cluster. Then, if the maximum inter-cluster distance of all clusters is less than the defined value ${\rm dis}^{\max}$, the adaptive cluster classification based on the K-means algorithm terminates. Otherwise, the number of clusters will increase by 1, and then the clusters will be classified again until the maximum intra-cluster distances of all clusters are less than the specific value  ${\rm dis}^{\max}$.

\begin{algorithm}[t] 
\caption{Adaptive Cluster Classification based on K-means Algorithm} 
\label{kmeans} 
\begin{algorithmic}[1]
	\STATE Initialize the iterative number $n = 1$ and the number of clusters $K_C^n$ = 2; 
 	\STATE Initialize the cluster based on the K-means algorithm and find the set of master APs.  
 	\STATE Calculate the maximum intra-cluster distance between the master AP and the slave AP of each cluster, denoted as ${\rm dis}_{k,n}^{\max}$, $\forall k \in \{1,\cdots, K_{C}^n\}$;
	\WHILE{($ \min\Big\{{\rm dis}_{1,n}^{\max},\cdots,{\rm dis}_{K_{C}^n,n}^{\max}\Big\} > {\rm dis}^{\max}$)} 
	\STATE Update $n = n + 1$, $K_{C}^n  = K_{C}^{n-1} + 1$; 
	\STATE Update the cluster classification ${\mathcal C}^{n} = \Big\{{\mathcal C}_1^{n} \cup  \cdots {\mathcal C}_{K_{C}^n}^{n}\Big\} $ with $K_{C}^n$, by using the K-means algorithm;
	\STATE Find the set of master APs by using (\ref{masterap});
	\STATE Calculate the maximum intra-cluster distance between the master AP and the slave AP of each cluster, denoted as ${\rm dis}_{k,n}^{\max}$, $\forall k \in \{1,\cdots, K_{C}^n\}$;
	\ENDWHILE 
\end{algorithmic} 
\end{algorithm}


\subsection{Pilot Sharing Algorithm}
Based on the given cluster ${\mathcal C}_k$, $\forall k \in \{1,\cdots, K_C\}$, we jointly optimize the pilot overhead $\tau$ and the pilot sharing scheme, ${\cal P}_p$, $\forall p \in \{1,\cdots, \tau\}$, to minimize the estimation errors. Mathematically, minimizing the sum of CRBs can be expressed as
\begin{subequations}
	\label{subproblem}
	\begin{align}
		\mathop {\min }\limits_{{\tau}, {{{\cal P}_p},p \in \left\{ {1, \cdots ,{\tau }} \right\}}}& 	\sum\limits_{k \in \left\{ {1,2, \cdots ,K_{C}} \right\}} {\sum\limits_{i \in \{{\cal C}_k \backslash j_k\}} {{{\rm{CRB}}_{i,j_k}(\Delta f_{i,j_k},\Delta \tau_{i,j_k})}} } \notag \\
		\text{s. t.} & \qquad {\rm CRB}_{i,j_k}(\Delta f_{i,j_k},\Delta \tau_{i,j_k}) \leq {\rm CRB}_{i,j_k}^{\rm Req}, \notag\\
		& \qquad \qquad j_k \in {\mathcal C}_k,\forall i \in \{{\mathcal C}_k \backslash j_k\}, k \in \{1,\cdots, K_{C}\}, \label{subconstrain1} \\
		& \qquad \qquad \tau \leq L_S - 2K_C. \label{subconstrain2}
	\end{align}
\end{subequations}
As can be seen from (\ref{subproblem}), the optimal solution can be obtained if all slave APs are allocated unique pilot sequences. However, this orthogonal pilot strategy is impractical owing to significant pilot overhead. To address this issue, we devise a pilot-sharing scheme by searching for the optimal solution with the given clusters.

In the following, we first simplify the pilot-sharing scheme by transforming it into a graph coloring problem. Specifically, two rules are defined to share the pilot sequence. For the first rule, to reduce the pilot contamination, the slave APs in the common cluster cannot share the same pilot sequence, which can be formulated as
\begin{equation}
	\label{1strule}
	{\bf s}_{p}^H {\bf s}_{p'} = 0, \quad \forall i^p  \quad {\rm and} \quad \forall i^{p'} \in {\mathcal C}_{k}, \quad \forall k \in\{1,\cdots, K_C\}, \quad {\rm and}\quad i^p \ne i^{p'},
\end{equation}
where ${\bf s}_{p} \in {\mathbb C}^{N \times 1}$ is the orthogonal pilot sequence allocated to the $i^p$-th slave AP. By denoting ${\cal S} = {\cal S}_1 \cup {\cal S}_2 \cdots {\cal S}_{K_C}$, we define a binary matrix $\bf B \in {\mathbb N}^{|\cal S| \times |\cal S|}$, where $|\cal S|$ denotes the number of element in ${\cal S}$, to indicate whether the $i$-th device is a potential candidate for sharing the common pilot sequence with the $i$-th device. The $i$-th row and $i'$-th column element of the matrix $\bf B$ is given by
\begin{equation}
	\setlength\abovedisplayskip{5pt}
	\setlength\belowdisplayskip{5pt}
	\label{bkk}
	{b_{i,i'}} = \left\{ {\begin{array}{*{20}{c}}
			{1,}& i \cup i' \notin {{\cal C}_{k,}}\forall k \in \left\{ {1, \cdots ,{K_C}} \right\}, {\text{and}} \quad i \ne i'\\
			{0,}&{{\text{otherwise,}}}
	\end{array}} \right.
\end{equation}
$b_{i,i'} = 1$ means that the $i$-th slave AP and the $i'$-th slave AP are in a common cluster, and thus these two APs should be allocated unique orthogonal pilot sequences. By contrast, $b_{i,i'} = 0$ denotes that the $i'$-th slave and $i$-th slave AP belong to different clusters, and it is possible to share the common pilot sequence between the $i$-th slave AP and the $i'$-th slave AP.

The second rule is to allocate the pilot sequence in a fair manner. Specifically, the $p$-th pilot sequence can only be reused no more than $N^{\max}$ times, which can be written as
\begin{equation}
	\label{2ndrule}
	|{\mathcal P}_p| \leq N^{\max}, \forall p \in \{1,\cdots, \tau\}.
\end{equation}
In this way, the worst-case scenario where a single pilot sequence is reused by too many devices can be avoided. 

\begin{figure*}
	\centering

	\subfigure[Undirected graph based on given clusters, , where S-AP and M-AP represent the slave AP and master AP, respecively.]{
		\begin{minipage}[t]{0.3\linewidth}
			\centering
			\includegraphics[width=2in]{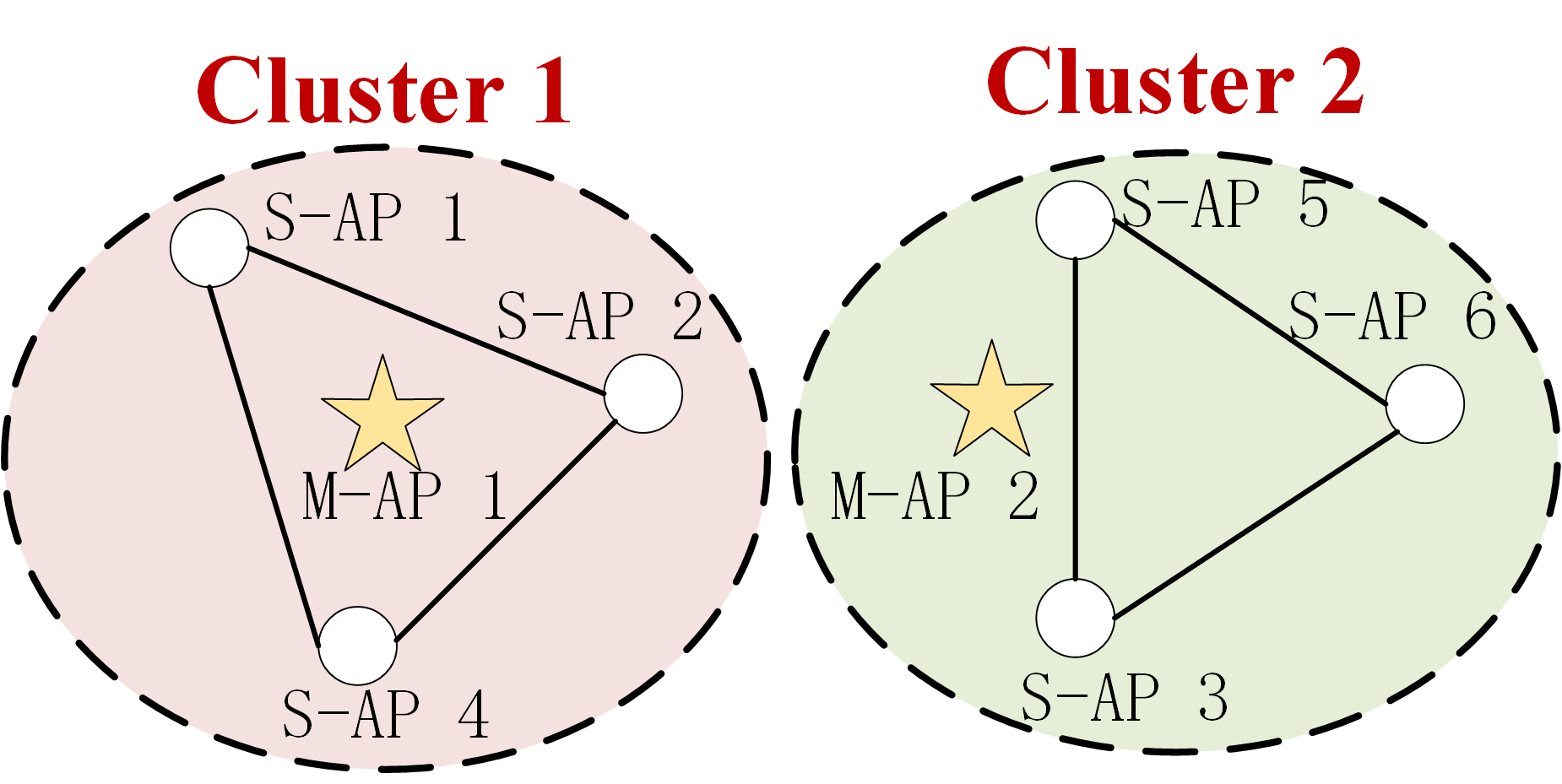}\hspace{10mm}
		\end{minipage}}
	\quad
	\subfigure[Pilot assignment based on the Dsatur algorithm, where the same color indicates that the same pilot sequence is used.]{\begin{minipage}[t]{0.3\linewidth}
			\centering
			\includegraphics[width=2in]{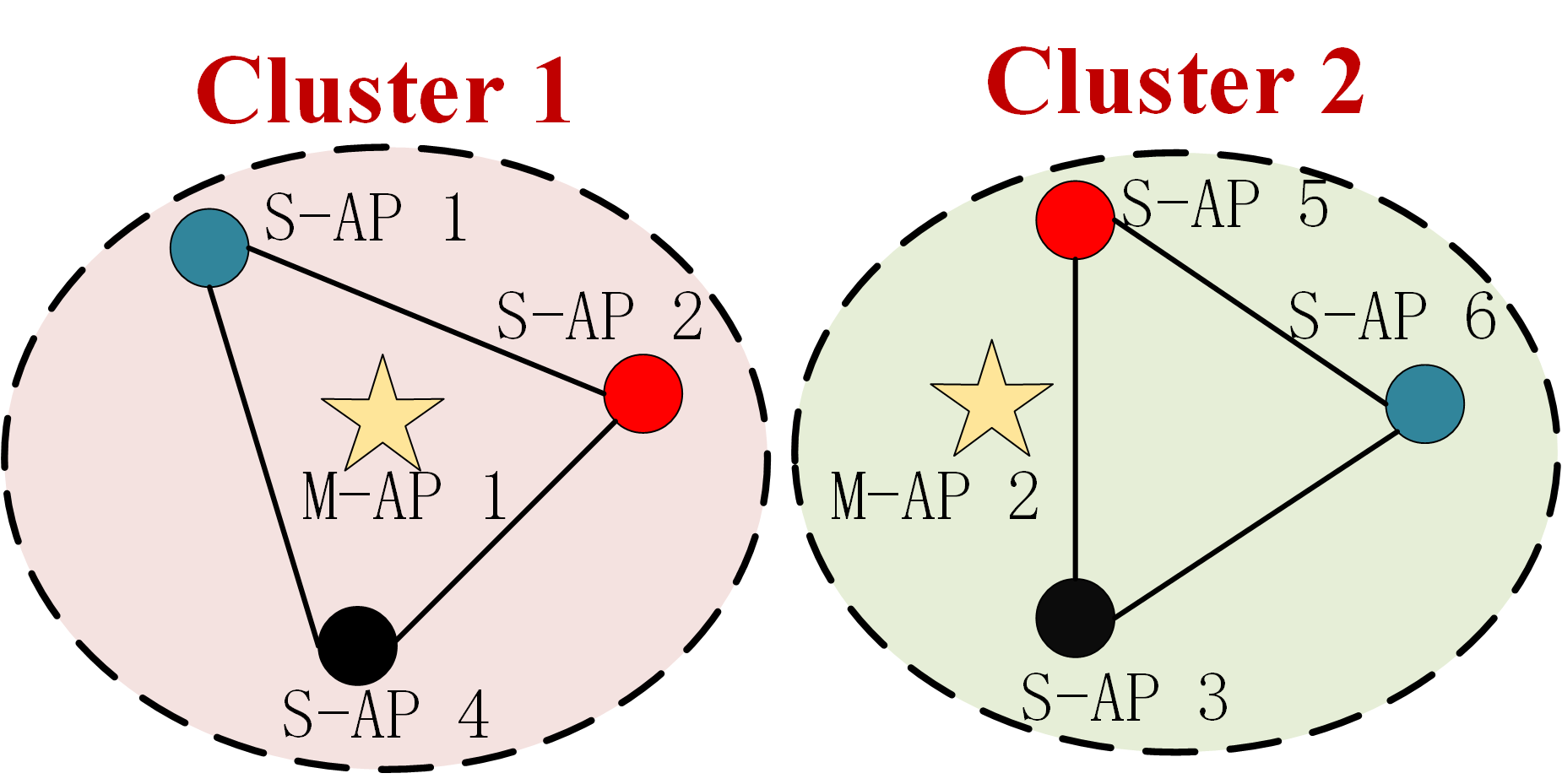}\hspace{10mm}
		\end{minipage}}
	\quad
	\subfigure[Pilot allocation after combining the Dsatur algorithm with the swap-matching operation.]{\begin{minipage}[t]{0.3\linewidth}
			\centering
			\includegraphics[width=2in]{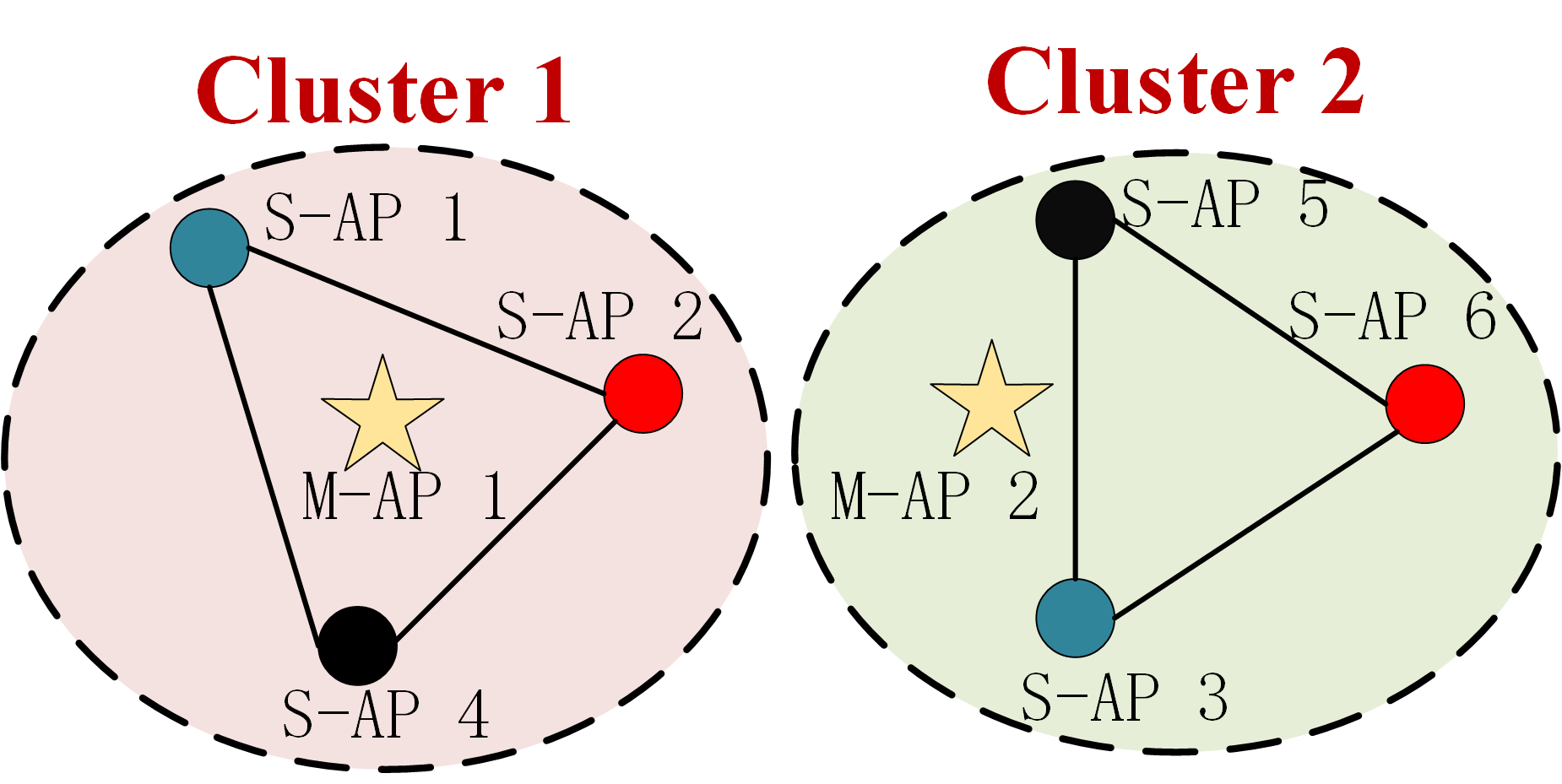} \hspace{10mm}
	\end{minipage}}
	\caption{Undirected graph and pilot allocation schemes.}
	\label{cluster}
\end{figure*}

By using the above two rules, allocating the pilot sequence can be transformed into the graph coloring problem, which can be effectively solved by using the Dsatur algorithm \cite{pan2018joint}. As depicted in Fig. \ref{cluster}, it is worth noting that a pilot allocation strategy based on the Dsatur algorithm can follow the above two rules by using the lowest pilot overhead. However, this algorithm did not consider the estimation errors. For example, based on the result given in (\ref{FSU}), the synchronization performance based on the pilot allocation strategy given in Fig. \ref{cluster} (b) is not optimal, as slave AP 5 and slave AP 2 that share the same pilot sequence are in close proximity, which leads to severe pilot contamination and poor synchronization performance.
Therefore, the objective function given in (\ref{subproblem}) cannot be minimized by simply using the Dsatur algorithm. To address this issue, we introduce the switch matching and two-sided exchange into our pilot sharing scheme.

In the following, we introduce the concept of the swap-matching operation.

{\bf Definition 1}: Given the pilot sharing scheme ${\cal P} = \{{\mathcal P}_1, \cdots, {\mathcal P}_{\tau}\}$ with $i \in {\mathcal P}_n$, $i' \in {\mathcal P}_m$, and $i \notin {\mathcal P}_m$, $i' \notin {\mathcal P}_n$, a swap-matching $\mu_{i,i'}$ is defined by the function ${\mathcal P}_n(\mu_{i,i'}) = {\mathcal P}_n \backslash i \cup  i' $ and ${\mathcal P}_m(\mu_{i,i'})  = {\mathcal P}_m \backslash i' \cup i$, and the exchanged pilot sharing scheme is defined as ${\cal P}_{\mu_{i,i'}} = \{{\mathcal P}_1, \cdots,{\mathcal P}_n(\mu_{i,i'}),{\mathcal P}_m(\mu_{i,i'}), {\mathcal P}_{\tau}\}$.

To be more specific, the swap-matching means a matching generated by swap operation \footnote{The swap operation is a two-sided version of the ``exchange" considered in \cite{bodine2011peer}.}, in which two slave APs using different pilot sequences switch their pilot sequences while keeping all other pilot allocation assignments the same. 

However, considering the previously defined two rules, the swap operation may not be allowed by all slave APs. Specifically, after this swap-matching operation, if the common pilot sequence is shared by the slave APs in the same cluster, then this operation is not permitted, which is detailed in the following.

{\bf Definition 2}: Given the pilot sharing scheme $\{{\mathcal P}_1, \cdots, {\mathcal P}_{\tau}\}$ with $i \in {\mathcal P}_n$, $i' \in {\mathcal P}_m$, and $i \notin {\mathcal P}_m$, $i' \notin {\mathcal P}_n$, a swap-blocking pair $(i,i')$ satisfies that:

(1): $\exists a,b \in {\mathcal P}_m(\mu_{i,i'})$ or $\exists a,b \in {\mathcal P}_n(\mu_{i,i'})$, $a \cup b \in {\cal C}_k$, $\forall k \in \{1,\cdots, K_C\}$.

Compared to the conventional swap-blocking pair, we do not care whether this swap-matching operation, in which all slave APs are involved, is two-sided exchange-stable. This is due to the fact that the two-sided exchange-stable match can only converge to the locally optimal pilot sharing scheme \footnote{Similar to the extremum of a multivariate function, it is impractical to obtain the globally optimal solution along the direction of one variable's partial derivative while holding the other variables constant. Therefore, given the other pilot sharing scheme, the swap-matching operation can only maximize its own benefits while ignoring the other slave APs' synchronization performance.}. To search the globally optimal matching with the given pilot overhead $\tau$, a swap-matching operation based on a simulated annealing method is proposed. Specifically, we will determine whether to update the pilot sharing scheme with a probability $P_a$, which is given by
\begin{equation}
	\label{probability}
	P_a = \frac{1}{1 + e^{-T \times [ {\rm Obj}({\cal P}_{\mu_{i,i'}})-{\rm Obj}({\cal P}_{\mu})  ]} },
\end{equation}
where $T$ is a probability parameter. ${\rm Obj}({\cal P}_{\mu_{i,i'}})$ is denoted as the sum of SINR based on the given pilot sharing scheme ${\cal P}_{\mu_{i,i'}}$, which can be written as
\begin{equation}
	\label{sumSINR}
	{\rm{Obj}}({\cal P}{_{{\mu _{i,i'}}}}) = \left\{ {\begin{array}{*{20}{c}}
			{\sum\limits_{k \in \left\{ {1, \cdots ,{K_C}} \right\}} {\sum\limits_{i_1 \in \left\{ {{{\cal C}_k}\backslash {j_k}} \right\}} {{\rm{SIN}}{{\rm{R}}_{i_1,{j_k}}},} } }&{{\rm{SIN}}{{\rm{R}}_{i_1,{j_k}}} \ge {\rm{SINR}}_{i_1,{j_k}}^{{\rm{Req}}}}\\
			{0,}&{{\rm{SIN}}{{\rm{R}}_{i_1,{j_k}}} < {\rm{SINR}}_{i_1,{j_k}}^{{\rm{Req}}}}.
	\end{array}} \right.
\end{equation}
As can be seen from the probability in (\ref{probability}), if the swapped pilot-sharing scheme is better than the previous one, then there is a high probability that the exchanged scheme will be accepted. Conversely, if any AP cannot satisfy the synchronization requirements, then the pilot allocation scheme may hold the same with high probability.

\begin{algorithm}[t] 
	\caption{Pilot-Sharing based on Swap-Matching Algorithm} 
	\label{pilotsharing} 
	\begin{algorithmic}[1]
		\STATE Initialize the first iteration number $n_1 = 1$, the maximum objective function ${\rm Obj}_{\max} = 0$, ${\rm Flag } = 1$, and the probability parameter $T$ is 0.01; 
		\STATE Initialize the maximum pilot overhead $\tau^{(n_1)} = \max \{{|{\cal S}_1|},\cdots,{|{\cal S}_{K_C}|}\}$, and the maximum number for reusing pilot sequences $N^{(n_1)}_{\max} = \left\lceil {\frac{{\left| {{{\cal S}_1} \cup {{\cal S}_1} \cdots {{\cal S}_{{K_{{C}}}}}} \right|}}{{\tau^{(n_1)}}}} \right\rceil $; 
		\STATE Use the Dastur algorithm to allocate the pilot sequences, denoted as $ {\cal P}^{(n_1)} = \{{\cal P}_1, \cdots, {\cal P}_{\tau^{(n_1)}}\}$;
		\WHILE{(${\rm Flag} == 1$)}
		\IF{($\tau^{(n_1)} > L_S - 2K_C$)}
		\STATE ${\rm Flag } = 0$;
		\ELSE
		\STATE Set the second iteration number $n_2$ to 1 and initialize the pilot sharing scheme as ${\cal P}^{(n_1,n_2)} = {\cal P}^{(n_1)}$;
		\WHILE{$n_2 \leq N_2^{\max} $}
		\STATE Update $n_2 = n_2 + 1$, randomly choose the proper slave APs and pilot sequence based on the two rules given in (\ref{1strule}) and (\ref{2ndrule}), and denote the switched pilot sharing scheme as ${\cal P}^{{\rm sw}}$; 
		\STATE Calculate the objective functions for both two pilot sharing schemes, ${\cal P}^{(n_1,n_2-1)}$ and ${\cal P}^{{\rm sw}}$, which are denoted as ${\rm Obj}^{(n_1,n_2-1)}$ and ${\rm Obj}^{{\rm sw}}$, and update pilot sharing scheme ${\cal P}^{(n_1,n_2)}$ by executing the swap-matching operation with probability $P_a$;
		\IF{(${\rm Obj}^{{\rm sw}} > {\rm Obj}^{(n_1,n_2-1)}$)} 
		\STATE Update the pilot sharing scheme ${\cal P} = {\cal P}^{{\rm sw}}$ and ${\rm Obj}_{\max} = {\rm Obj}^{{\rm sw}}$;
		\ELSE
		\STATE Update the pilot sharing scheme ${\cal P} = {\rm Obj}^{(n_1,n_2-1)}$ and ${\rm Obj}_{\max} = {\rm Obj}^{(n_1,n_2-1)}$;
		\ENDIF
		\ENDWHILE
		\STATE Update $n_1 =n_1 + 1$, $\tau_p^{(n_1)} = \tau^{(n_1 -1)} + 1$, $N^{(n_1)}_{\max} = \left\lceil {\frac{{\left| {{{\cal S}_1} \cup {{\cal S}_1} \cdots {{\cal S}_{{K_{{C}}}}}} \right|}}{{\tau^{(n_1)}}}} \right\rceil $;
		\STATE Randomly select ($N^{(n_1)}_{\max} - 1$) slave APs with low SINRs, replace their pilots with the $\tau^{(n_1)}$-th pilot sequence, and denote the new pilot sharing scheme as ${\cal P}^{(n_1)}$;
		\ENDIF
		\ENDWHILE
		
	\end{algorithmic} 
\end{algorithm}

In this way, we can keep tracking the optimal matching found in an iterative manner in a feasible region, i.e., $\tau \leq L_S - 2K_C$, even if the utility of the current matching is not locally optimal, which is detailed in Algorithm \ref{pilotsharing}. Specifically, based on the coloring theory, in the $n_1$-th iteration, the pilot overhead $\tau^{(n_1)}$ is initialized as the maximum number of slave APs in one cluster, and the maximum number for reusing one pilot sequence is determined by the number of slave APs and the pilot overhead, which is defined as
\begin{equation}
	\label{maxre}
	N^{(n_1)}_{\max} = \left\lceil {\frac{{\left| {{{\cal S}_1} \cup {{\cal S}_1} \cdots {{\cal S}_{{K_{{C}}}}}} \right|}}{{\tau^{(n_1)}}}} \right\rceil,
\end{equation}
where $\left\lceil \cdot \right\rceil$ means the ceiling function operation. After adopting the Dastur algorithm, we will check whether constraint (\ref{subconstrain2}) holds or not. If the total overhead is larger than $L_S$, Algorithm 2 will terminate. Conversely, if the total overhead is less than $L_S$, we begin to search the optimal solution with given pilot overhead. Particularly, if the total overhead is less than $L_S$, we begin to search the optimal solution. With the given pilot overhead $\tau^{(n_1)}$, the swap-matching operation is repeated until the number of iterations $n_2$ exceeds a specified number $N_2^{\max}$. Furthermore, after each swap-matching operation, the pilot sharing scheme with better performance is updated. In this way, the optimal pilot sharing scheme with pilot overhead $\tau^{(n_1)}$ can be found based on the synchronization requirements. Then, to guarantee that all pilot sequences are fully utilized, the available pilot sequence will be allocated to the slave APs that cause severe pilot contamination. Particularly, we update the number of iterations by $n_1 = n_1 + 1$, the pilot overhead by $\tau^{(n_1)} = \tau^{(n_1 - 1)}  + 1$, and the number for reusing pilot sequence by $N^{(n_1)}_{\max} = \left\lceil {\frac{{\left| {{{\cal S}_1} \cup {{\cal S}_1} \cdots {{\cal S}_{{K_{{C}}}}}} \right|}}{{\tau^{(n_1)}}}} \right\rceil $. Then, we randomly select ($N_{\max}^{(n_1)} - 1$) slave APs with low SINRs, denoting their indices as $\{i^{p_1},\cdots,i^{p_{N_{\max}^{(n_1)} - 1}}\}$ and their pilot sequences as $\{{p_1},\cdots, {p_{N_{\max}^{(n_1)} - 1}}\}$, ${p_m} \neq {p_{m'}}$ if $m \neq m'$. Furthermore, to follow the first rule, for given any two selected APs, $i^{p_m}$ and $i^{p_{m'}}$, $\forall m,m' \in \{1,\cdots, {N_{\max}^{(n_1)} - 1}\}$, $m \neq m'$, two APs are not in the common cluster, which is given by 
\begin{equation}
	\setlength\abovedisplayskip{5pt}
	\setlength\belowdisplayskip{5pt}
	\label{selectedAP}
	i^{p_m} \cup i^{p_m'} \notin {\cal C}_k, \forall k \in \{1,\cdots, K_C\}.
\end{equation}
After that, the new pilot sequence is assigned to the selected slave APs, and the pilot sharing scheme in the $n_1$-th iteration is updated, which is given by
\begin{equation}
	\setlength\abovedisplayskip{5pt}
	\setlength\belowdisplayskip{5pt}
	\label{randomly}
	\begin{split}
		{{\cal P}^{(n_1)}_{{p_m}}} &\!=\! \left\{ \!{{{\cal P}^{(n_1 - 1)}_{{p_m}}}\backslash i^{p_m}} \!\right\},\forall m \in \left\{ {1, \cdots ,{N^{({n_1})}_{{{\max }}}} - 1} \right\}, \\
		{{\cal P}^{(n_1)}_{\tau^{(n_1)}}} &=   \left\{ i^{p_1},\cdots, i^{p_{{N^{(n_1)}_{\max} - 1}}}\right\}.
	\end{split}
\end{equation}
Finally, based on the given clusters $\{{\cal C}_1, \cdots,{\cal C}_{K_C}\}$, the optimal pilot sharing scheme is obtained by searching the feasible solutions.

To find the optimal solution to Problem (\ref{optproblem}), the above two algorithms are executed iteratively and the optimal solution is reserved, which is detailed in Algorithm \ref{ab}.

\begin{algorithm}[t] 
	\caption{Resource Allocation for Problem (\ref{optproblem})} 
	\label{ab} 
	\begin{algorithmic}[1]
		\STATE Initialize iterative number $n = 0$ and the number of clusters $K_C^n = 2$ ;
		\STATE  Initialize the $\tau^n = 0$ and sum of CRBs as ${\rm CRBs}^{(n)} = +\infty $;
		\WHILE{($\tau^n + 2K^n_C \leq L_S$)} 
		\STATE Update $n = n + 1$, execute Algorithm \ref{kmeans} with given  $K_C^n$, and obtain the clusters $\{{\cal C}_1,\cdots,{\cal C}_{K_C^n}\}$;
		\STATE Execute Algorithm \ref{pilotsharing} to obtain the pilot allocation scheme in the $n$-th iteration, denoted as $\{{\cal P}_1,\cdots,{\cal P}_{\tau^n}\}$;
		\STATE Calculate the sum of CRBs, denoted as ${\rm CRBs}^{(n)}$;
		\STATE Update $K_C^{n+1} = K_C^{n} + 1$;
		\IF{(${\rm CRBs}^{(n)} < {\rm CRBs}^{(n-1)}$)}
		\STATE Update cluster classification ${\cal C} = \{{\cal C}_1,\cdots,{\cal C}_{K_C^n}\}$ and pilot allocation ${\cal P} = \{{\cal P}_1,\cdots,{\cal P}_{\tau^n}\}$;
		\ENDIF
		\ENDWHILE 
	\end{algorithmic} 
\end{algorithm}

\subsection{Algorithm Analysis}
In this subsection, the complexity of our proposed algorithm is analyzed. For the cluster classification algorithm, the average complexity is $O(L_C \times K \times I_c)$, where $I_c$ is the number of iterations. For the pilot-sharing scheme, we need to search the optimal solution subject to the pilot overhead budget. Furthermore, for a given pilot overhead, the swap-matching operation needs to be executed $N_2^{\max}$ times. The complexity of the pilot-sharing scheme is $O(N_2^{\max} \times I_{\tau} )$, where $I_{\tau} $ is the number of total iterations. Therefore, the complexity of our proposed algorithm lies in the number of both iterations of cluster classification and swap-matching operations.

\section{Simulation Results}
In this section, the performance of our proposed pilot allocation strategy is numerically evaluated and discussed.

\subsection{Simulation Parameters}
There are $K$ APs that are uniformly positioned constellation points in a smart factory of size $0.15 \times 0.15$ ${\text{km}}^2$. The large-scale fading factors are based on the Hata-COST231 propagation model of \cite{Ngo2017}, where the height of each AP is 15 m. Specifically, the three-slope path loss (dB) can be expressed as
\begin{equation}
	\setlength\abovedisplayskip{5pt}
	\setlength\belowdisplayskip{5pt}
	\label{channel_model}
	{\rm{P}}{{\rm{L}}_{m,k}} = \left\{ {\begin{array}{*{20}{l}}
			{\begin{array}{*{20}{l}}
					{L_{\rm{loss}} + 35{{\log }_{10}}\left( {{d_{m,k}}} \right),}\\
					{L_{\rm{loss}} + 15{{\log }_{10}}\left( {{d_1}} \right) + 20{{\log }_{10}}\left( {{d_{m,k}}} \right),}\\
					{L_{\rm{loss}} + 15{{\log }_{10}}\left( {{d_1}} \right) + 20{{\log }_{10}}\left( {{d_0}} \right),}
			\end{array}}&{\begin{array}{*{20}{c}}
					{{d_{m,k}} > {d_1}}\\
					{{d_0} < {d_{m,k}} \le {d_1}},\\
					{{d_{m,k}} \le {d_0}}
			\end{array}}
	\end{array}} \right.
\end{equation}
where $d_{m,k}$ is the distance between the $m$-th AP and the $k$-th AP, $L_{\rm{loss}}$ is a constant of 140.7 (dB), while $d_0$ and $d_1$ are 0.01 km and 0.05 km, respectively. Furthermore, each AP's transmission power is 1 W.

\subsection{Derivation and Monte-Carlo Simulations}

\begin{figure}
	\centering
	\includegraphics[width=4.5 in]{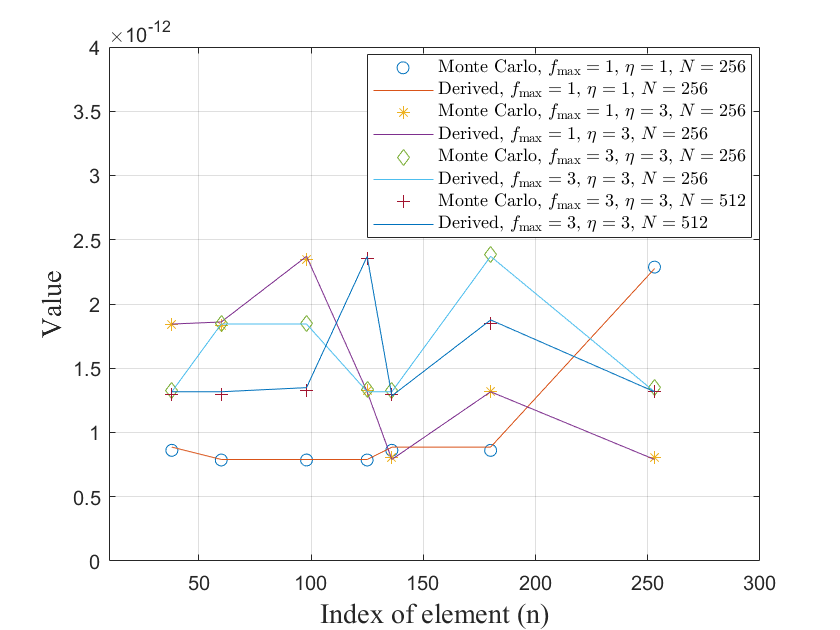}
	\caption{$[{{{\bf{D}}_{i,{j_k}}}}]_{n,n}$ based on Monte-Carlo simulation and derived result.}
	\label{element}
\end{figure}

\begin{figure}
	\centering
	\includegraphics[width=4.5 in]{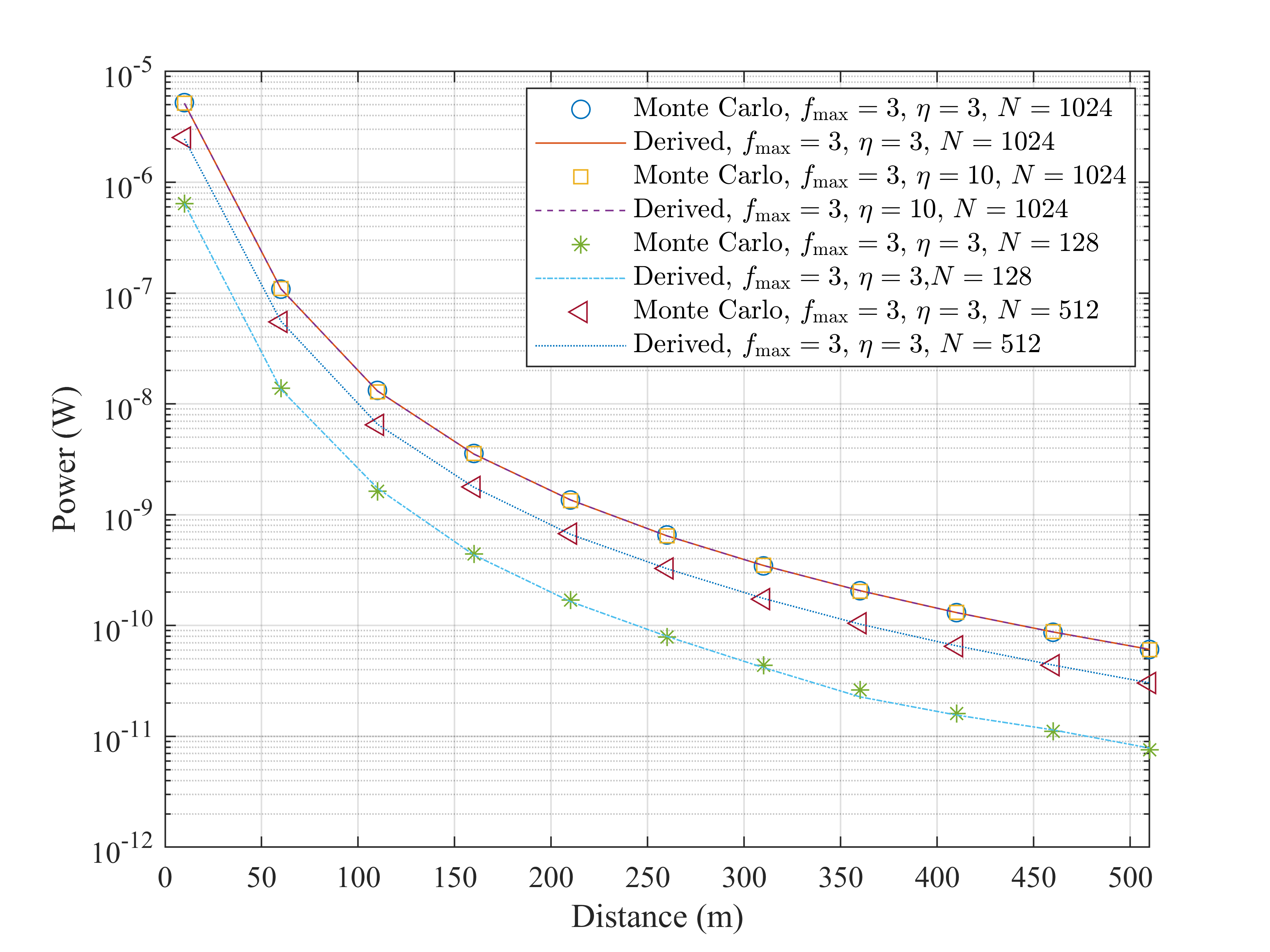}
	\caption{Power of Pilot Contamination based on Monte-Carlo simulation and derived result.}
	\label{MCVSD}
\end{figure}
To check whether our derivations can provide a more convenient expression for the pilot sharing scheme, we first check the $[n,n]$-th element of ${\bf D}_{i,j_k}$ based on Monte-Carlo simulations and derived result. Fig. \ref{element} depicts the randomly chosen elements of ${\bf D}_{i,j_k}$ with different CFOs, TOs, and pilot lengths. Monte-Carlo simulation is consistent with our derived closed-form expression, which demonstrates the accuracy of our derivations. 

To investigate the relationship between the distance-based large-scale fading factor and the power of pilot contamination, we depict the power of derived pilot contamination and that based on Monte-Carlo simulations. Fig. \ref{MCVSD} shows the power with a pilot sequence $\bf s$ formed by various lengths $N$ in the BPSK constellation. Obviously, our derived results are consistent with Monte-Carlo simulations, demonstrating that our derived result can provide an explicit expression for the latter pilot-sharing scheme. Furthermore, we observe that the power of pilot contamination increases with the pilot length $N$. This is due to the fact that the power is equal to the trace of the matrix, which increases with the pilot length $N$. More importantly, it is worth noting that the power of pilot contamination has no significant relationship between the CFO and TO but significantly decreases with the increased distance, which demonstrates the effectiveness of our derivations.

\subsection{Cramer-Rao Bound and ML Estimation}
\begin{figure}
	\centering
	\includegraphics[width=4.5 in]{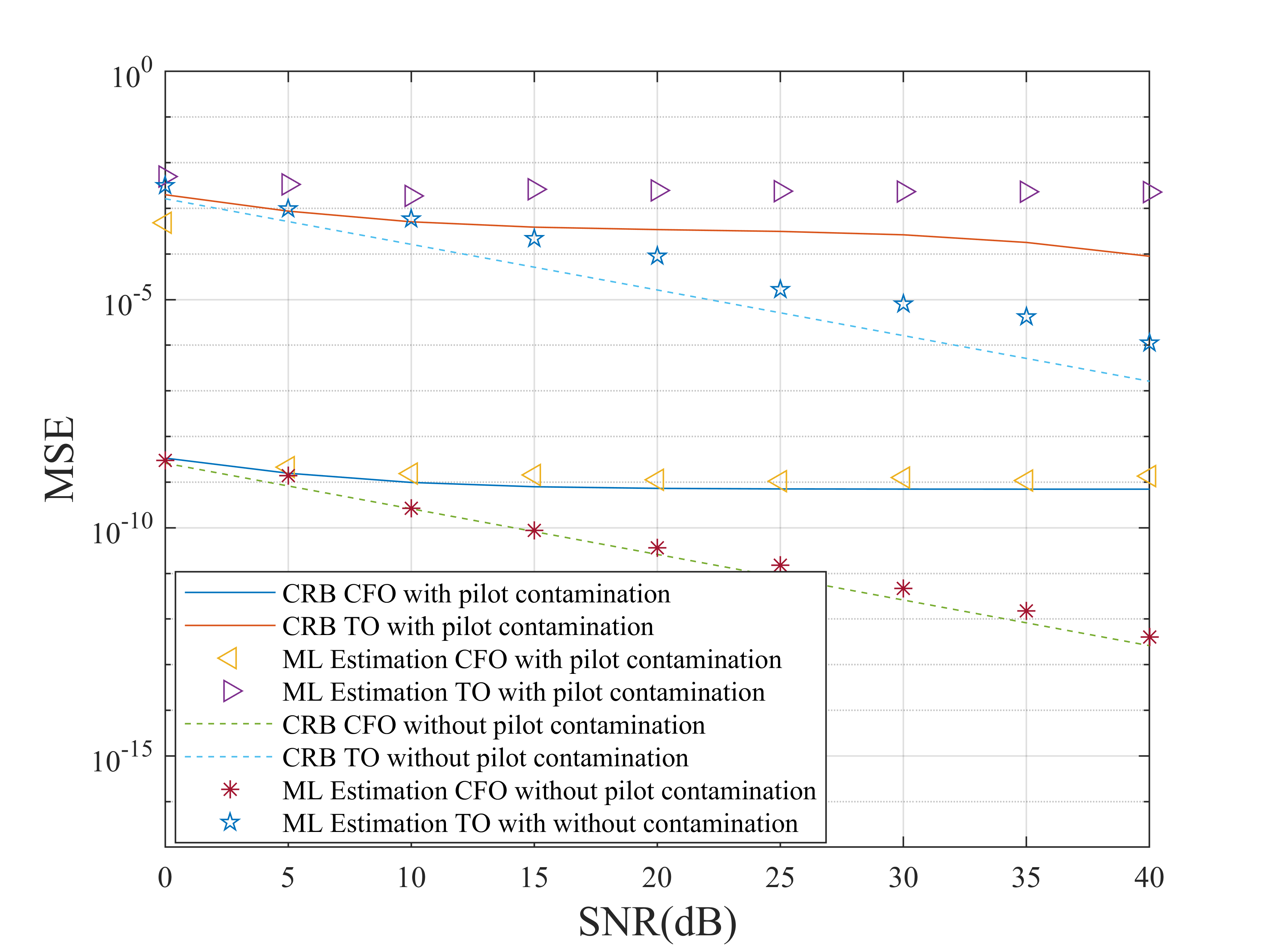}
	\caption{CRB and ML estimation versus SNRs with path delay [0, 0.7049$T_s$, 2.1230$T_s$, 2.7063$T_s$], path gain [1, 0.8443,0.4119,0.3223], and large-scale fading factor 5.16 $\times 10^{-10}$.}
	\label{CRLBCFOTO}
\end{figure}

To verify our derivations, we first check the theoretical lower bound, i.e., CRB, and the MSE based on ML estimation. Fig. \ref{CRLBCFOTO} depicts the synchronization performance with various SNRs. Furthermore, to obtain the inverse matrix, the large-scale fading factor of each path is normalized by dividing the maximum large-scale fading factor from the slave AP to master AP. Apparently, the CRBs and MSEs of ML estimation decrease with the increased SNR, especially for the results without pilot contamination. This phenomenon arises due to the diminished interference between the desired signal and noise when the SNR is high, resulting in better synchronization performance. However, owing to the pilot sharing, the lower bound based on the pilot contamination no longer linearly decreases with SNR, but instead experiences a flat trend. This is due to the fact that as the noise power decreases, the main factor affecting the estimation performance is no longer the noise, but the pilot contamination. Besides, compared to the case without pilot contamination, the performance gap relying on pilot sharing between ML estimation and CRB is enlarged, owing to the significant effects of pilot contamination.


\subsection{Performance of the Proposed Algorithm}

\begin{figure}
	\centering
	\includegraphics[width=4.5 in]{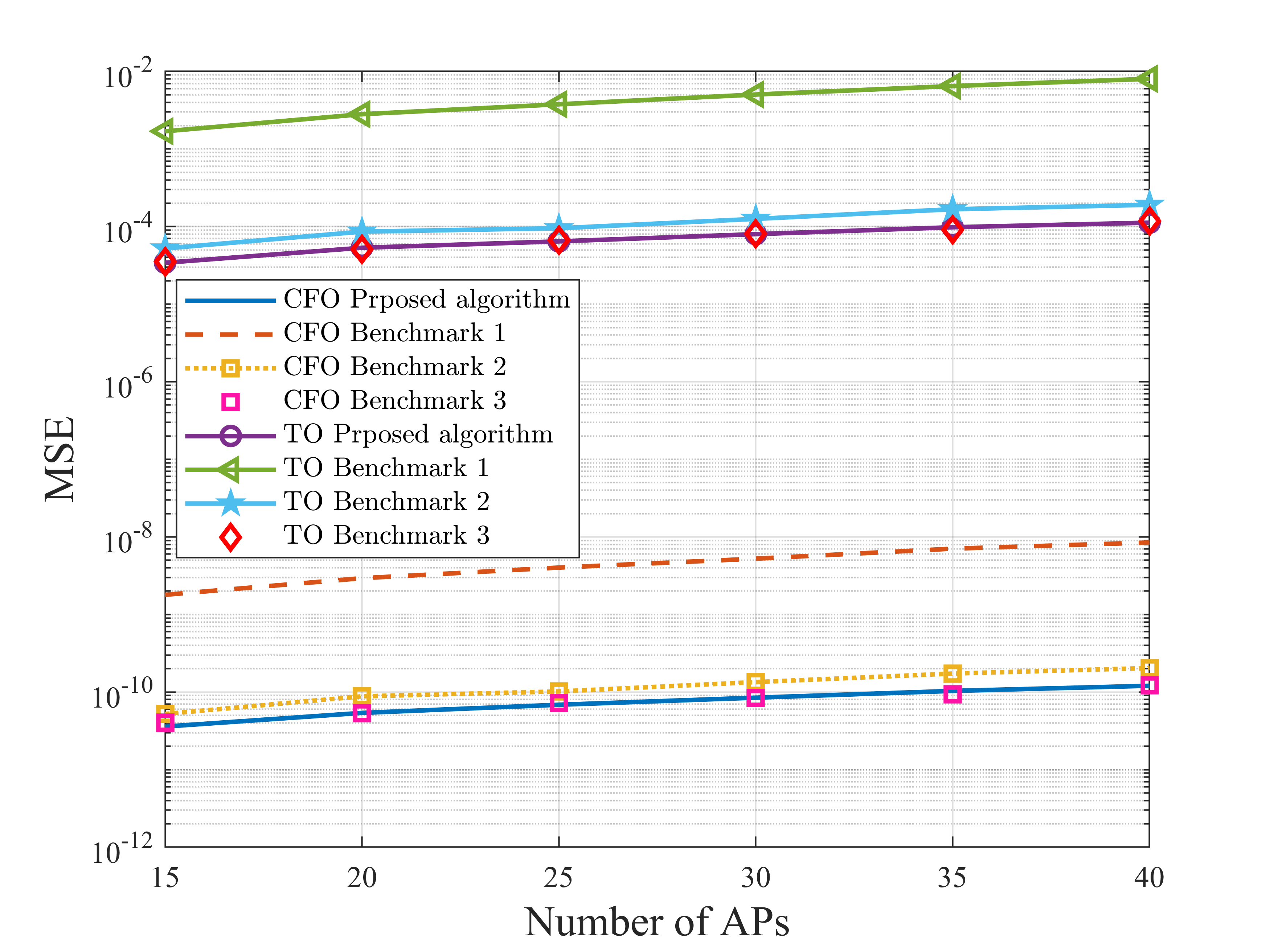}
	\caption{Sum of CRBs versus the number of APs with $N = 256$, $\eta = 3$, $f_{\max} = 3$, ${\rm SINR}^{\min} = 15$ {\rm dB}.}
	\label{adaptiveCRLB}
\end{figure}

To investigate the impact of clustering on the synchronization performance, we assume that the $K$ APs are randomly distributed in this area and all slave APs are allocated to a unique pilot sequence. Furthermore, to illustrate the performance of our adaptive cluster classification, we also compare our approach with the following three benchmarks that may violate the constraints in (\ref{optproblem}):

{\bf Benchmark 1}: All slave APs transmit pilot sequences directly to the CPU, whose 2-dimensional location is [0,0].

{\bf Benchmark 2}: The clusters are classified by the K-means algorithm while the master APs are randomly chosen in each cluster.

{\bf Benchmark 3}: The clusters are classified relying on the hierarchy clustering algorithm while the master APs are chosen based on our criteria given in (\ref{masterap}).

By averaging over 100 simulation results where the path delays and path gain coefficients are randomly generated, Fig. \ref{adaptiveCRLB} illustrates the sum of CRBs of all slave APs. Obviously, the synchronization performance slightly increases with the number of APs. Furthermore, it is worth noting that our approach is superior to all benchmarks. This is due to the fact that the cluster classification could shorten the distance between master and slave APs, thereby improving the path gain and enhancing the estimation performance. Additionally, in each cluster, the synchronization performance is further improved by reducing the distance between master APs and slave APs.
\begin{figure*}
	\centering
	\subfigure[Synchronization performance.]{
		\includegraphics[width=2.75in]{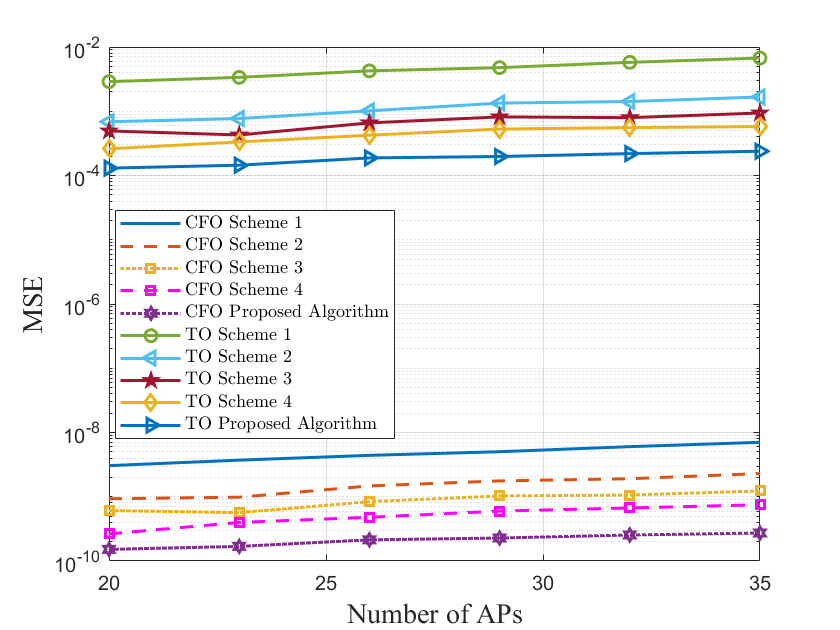}}\hspace{10mm}
	\subfigure[Synhchronization performance.]{
		\includegraphics[width=2.75in]{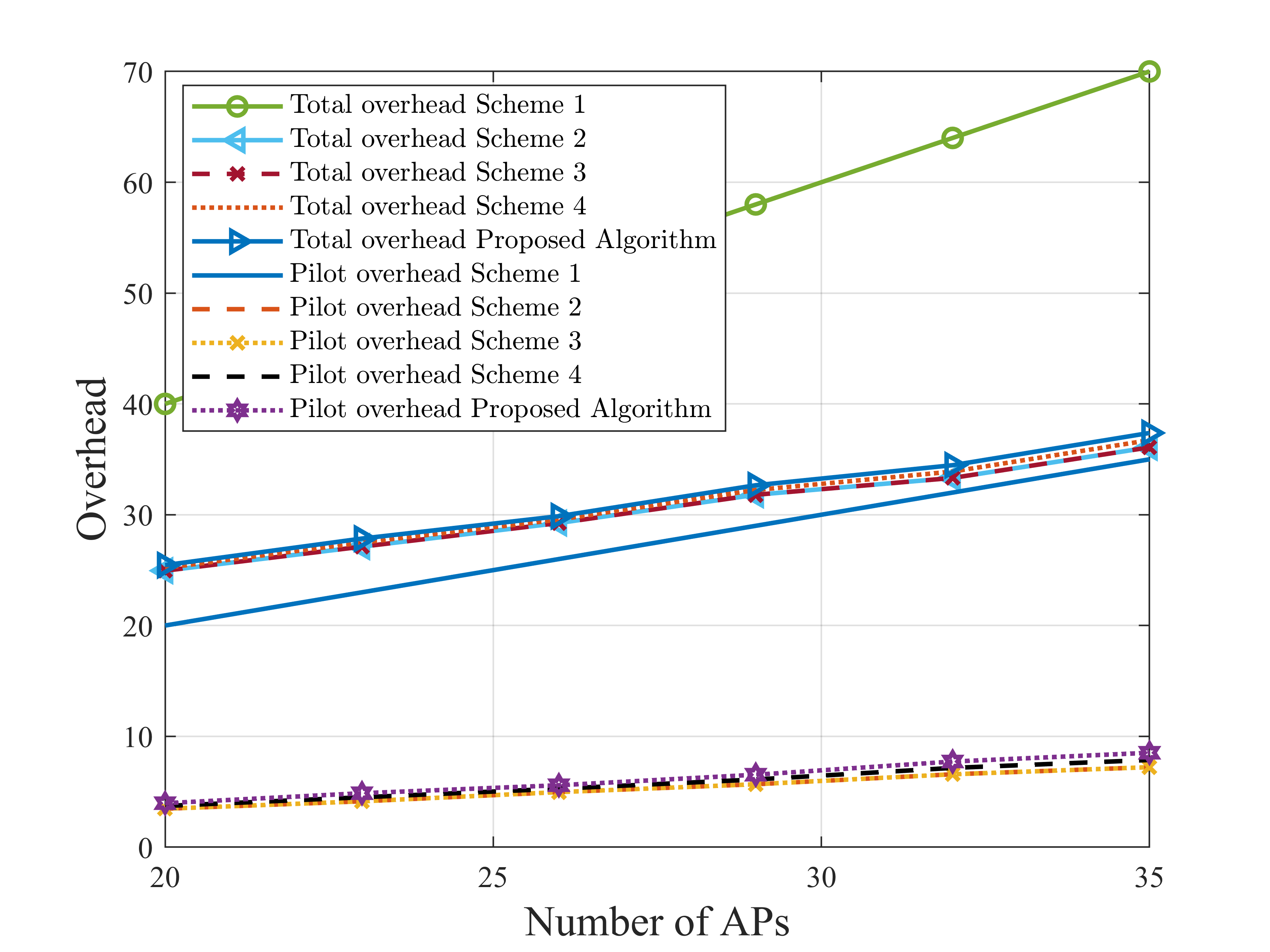}}
	\caption{Sum of CRBs versus the number of APs with $N = 256$, $\eta = 3$, $f_{\max} = 3$, ${\rm SINR}^{\min} = 15$ dB, $L_S = [25, 28, 31, 34, 37, 40]$.}
	\label{proposedalgorhtm}
\end{figure*}



To further investigate the performance of our pilot sharing algorithm, we also compare our algorithm with the following pilot allocation schemes.

{\bf Scheme 1}: All salve APs transmit orthogonal pilot sequences directly to the CPU, whose 2-dimensional location is [0,0].

{\bf Scheme 2}: The clusters are classified by the K-means algorithm while the master APs are randomly chosen in each cluster, and the pilot sharing scheme is obtained by using the Dastur algorithm.

{\bf Scheme 3}: The clusters are classified by using our adaptive cluster algorithm, and the pilot sharing scheme is obtained by using the Dastur algorithm.

{\bf Scheme 4}: The clusters are classified relying on Benchmark 3 and the pilot sharing scheme is obtained by using the improved Dastur algorithm in \cite{peng2023resource}.

Fig. \ref{proposedalgorhtm}(a) illustrates the averaged sum of CRBs under the case of various numbers of APs, while  Fig. \ref{proposedalgorhtm}(b) shows the averaging overhead of each scheme. 
As can be seen from Fig. \ref{proposedalgorhtm}(a), the synchronization performance slightly increases with the number of APs. For Scheme 1, owing to the lower path gain, the estimation errors are unable to meet the synchronization requirements. Even though sharing pilots causes pilot contamination between APs, the distance among APs that use the common pilot sequence is maximized to meet the synchronization requirements. Furthermore, benefiting from our proposed algorithm, the estimation errors are comparatively stable in the demand range. As for the overhead shown in Fig. \ref{proposedalgorhtm}(b), Scheme 3 has the lowest pilot overhead, as it focuses on the minimal number of colors (i.e., pilot overhead). Similarly, Scheme 4 can only obtain the locally optimal solution with lower pilot overhead compared to our proposed algorithm. However, in the cell-free mMIMO system, we aim to strike a good balance between the synchronization overhead and performance.

\section{Conclusions}
To mitigate the impacts of synchronization errors on cooperative ISAC, we investigated the impact of synchronous pilot sharing on the estimation performance by deriving the CRB while considering the pilot contamination. Furthermore, a maximum likelihood algorithm was presented to estimate the CFO and TO among the pairing APs. Then, to minimize the sum of synchronization errors, a novel synchronization strategy based on the pilot-sharing scheme was devised by jointly optimizing the cluster classification, synchronization overhead, and pilot-sharing scheme. This NP-hard problem was simplified into two sub-problems, including cluster classification and the pilot sharing scheme. Specifically, the clusters were classified by combining the K-means algorithm with the minimum distance sum criterion, while the pilot sharing scheme was obtained using swap-matching operations. Simulation results validated the accuracy of our derivations and demonstrated the effectiveness of the proposed scheme over other pilot-pair algorithms.

\begin{appendices}
	\section{Proof of (\ref{model})}
	\label{Proof}
	The $[k',l]$-th element of ${{{\bf{S}}_p}{\bf{U}}(\Delta {\tau _{i,{j_k}}})}$ is given by
	\begin{equation}
		\label{SpU}
		\begin{split}
			{\left[ {{{\bf{S}}_p}{\bf{U}}(\Delta {\tau _{i,{j_k}}})} \right]_{k',l}} & = \sum\limits_{n = 1}^{M - N + 1} {{{\left[ {{{\bf{S}}_p}} \right]}_{k',n}}{{\left[ {{\bf{U}}(\Delta {\tau _{i,{j_k}}})} \right]}_{n,l}}} \\
			& = \sum\limits_{n = 1}^{M - N + 1} {{{\left[ {{{\bf{S}}_p}} \right]}_{k',n}}\nabla (n - 1 - \frac{{\Delta {\tau _{i,{j_k}}} + \varsigma _{i,{j_k}}^l}}{{{T_s}}})} \\
			& = \sum\limits_{n = 1}^{M - N + 1} {s_p^{k' - n}\nabla (n - 1 - \frac{{\Delta {\tau _{i,{j_k}}} + \varsigma _{i,{j_k}}^l}}{{{T_s}}})}, 
		\end{split}
	\end{equation}
where $s^{k'-n}_p$ is $0$ when $k'-n < 0$ and $k'-n > N-1$.
	
Based on the definition of $\nabla \left( t \right)$, we have
\begin{equation}
	\label{nabla}
	\nabla \left( t \right) = \left\{ {\begin{array}{*{20}{c}}
			{t,}&{0 < t \le 1}\\
			{2 - t,}&{1 < t \le 2}\\
			{0,}&{\rm otherwise}.
	\end{array}} \right.
\end{equation}

By substituting (\ref{nabla}) into (\ref{SpU}), we have 
\begin{equation}
{\left[ {{{\bf{S}}_p}{\bf{U}}(\Delta {\tau _{i,{j_k}}})} \right]_{k',l}} = s_p^{k' - {x_1}}\nabla ({x_1} - 1 - \frac{{\Delta {\tau _{i,{j_k}}} + \varsigma _{i,{j_k}}^l}}{{{T_s}}}) + s_p^{k' - {x_1} - 1}\nabla ({x_1} - \frac{{\Delta {\tau _{i,{j_k}}} + \varsigma _{i,{j_k}}^l}}{{{T_s}}}),
\end{equation}
where ${x_1}$ is $\left\lceil {1 + \frac{{\Delta {\tau _{i,{j_k}}} + \varsigma _{i,{j_k}}^l}}{{{T_s}}}} \right\rceil $.

Then, letting $k'-x_1 = n$, $k' \in \{1,\cdots, M\}$, we have 
\begin{equation}
	\label{m1}
	\begin{split}
		{\left[ {{{\bf{S}}_p}{\bf{U}}(\Delta {\tau _{i,{j_k}}})} \right]_{k',l}} &= s_p^n\nabla (k' - 1 - n - \frac{{\Delta {\tau _{i,{j_k}}} + \varsigma _{i,{j_k}}^l}}{{{T_s}}}) + s_p^{n - 1}\nabla (k' - n - \frac{{\Delta {\tau _{i,{j_k}}} + \varsigma _{i,{j_k}}^l}}{{{T_s}}})\\
	&= \sum\limits_{n = 0}^{N - 1} {s_p^n\nabla (k' - 1 - n - \frac{{\Delta {\tau _{i,{j_k}}} + \varsigma _{i,{j_k}}^l}}{{{T_s}}})}.
	\end{split}
\end{equation}

Finally, we substitute $m = k'-1$ into (\ref{m1}) and complete the proof of (\ref{model}) by combining the results of ${\bf F}(\Delta f_{i,j_k})$ and ${\bf h}_{i,j_k}$.
\end{appendices}

\bibliography{myref,ref/bibfile}

\end{document}